\documentclass[5p]{elsarticle}
\usepackage[utf8]{inputenc}
\usepackage{bm}
\usepackage{amsmath}
\usepackage{mathtools}
\usepackage[colorlinks=true, pdfstartview=FitV, linkcolor=black, citecolor= black, urlcolor= black]{hyperref}
\usepackage{footnpag}
\usepackage{gensymb}
\usepackage{multirow}
\usepackage{bm}
\usepackage[ruled,vlined]{algorithm2e}
\usepackage{multicol}
\usepackage{enumitem}

\biboptions{sort&compress}
\begin{document}

\begin{frontmatter}

\title{Autonomous Angles-Only Multitarget Tracking for Spacecraft Swarms \tnoteref{t1}}
\tnotetext[t1]{This research has been supported by the NASA Small Spacecraft Technology Program cooperative agreement number 80NSSC18M0058.}

\author[1]{Justin Kruger\corref{cor1}}
\ead{jjkruger@stanford.edu}
\author[1]{Simone D'Amico}
\ead{damicos@stanford.edu}

\cortext[cor1]{Corresponding author}

\address[1]{Department of Aeronautics and Astronautics, Stanford University, Stanford, CA, 94305, USA}

\begin{abstract}
This paper presents a new algorithm for autonomous multitarget tracking of resident space objects using optical angles-only measurements from a spaceborne observer. To enable autonomous angles-only navigation of spacecraft swarms, an observer must be able to identify and track multiple known or unknown target space objects in view, without reliance on a-priori relative orbit knowledge. Extremely high assignment precision is necessary with low measurement frequencies and limited computational resources. The new ‘Spacecraft Angles-only MUltitarget tracking System’ (SAMUS) algorithm has been developed to meet these objectives and constraints. It combines domain-specific modeling of target kinematics with multi-hypothesis techniques to autonomously track multiple unknown targets using only sequential camera images. A measurement transform ensures that target motion in the observer reference frame follows consistent parametric models; curve fitting is used to predict track behavior; and kinematically-derived track gating and scoring criteria are applied to improve the efficiency and accuracy of the multi-hypothesis approach. Monte-Carlo testing with high-fidelity simulations demonstrates close to 100\% measurement assignment precision and strong recall across a range of multi-spacecraft formations, in both near-circular and eccentric orbits. Tracking is maintained in the presence of eclipse periods, significant measurement noise, and partially known swarm maneuvers. A comparison to other tracking algorithms reveals strong advantages in precision, robustness and computation time, crucial for spaceborne angles-only navigation.
\end{abstract}

\begin{keyword}
navigation \sep multitarget tracking \sep distributed space systems \sep angles-only
\end{keyword}

\end{frontmatter}

\section{Introduction}
\label{intro}
\noindent Distributed space systems can offer many advantages over traditional monolithic spacecraft, including improved coverage, costs, scalability, flexibility and robustness \cite{intro1, intro2}. However, their navigation presents significant challenges, especially in the context of deep space missions aiming to navigate primarily autonomously using only on-board resources. For spacecraft swarms operating at separations of several kilometers to several thousand kilometers, a favorable solution is angles-only navigation, in which observer spacecraft obtain bearing angle measurements to targets using an on-board vision-based sensor (VBS). Cameras are advantageous as they are robust, low-cost, low-power sensors already present on most spacecraft. They possess high dynamic range capabilities and small form factors conducive to both accurate navigation and swarm miniaturization. Many distributed space system proposals therefore present angles-only navigation as a key aspect, with applications to distributed science \cite{intro4, marsao}, space situational awareness, deep space communications \cite{starling}, autonomous rendezvous \cite{argon, avanti}, and on-orbit servicing \cite{intro3, intro5}.
\par
A number of studies have explored angles-only navigation for spacecraft. Woffinden et al. \cite{ao1} and Gaias et al. \cite{ao2} discuss angles-only state estimation using linearized rectilinear relative motion and relative orbital elements (ROE) respectively. They conclude that the linearized angles-only navigation problem is not fully observable due to a lack of explicit range information and suggest conducting maneuvers to improve observability. This, however, is not ideal, as navigation and control then become coupled. Sullivan et al. \cite{josh, generalized} subsequently presented a maneuver-free procedure for angles-only navigation, by leveraging nonlinearities in the form of perturbed orbit dynamics and orbit curvature effects for improved state estimation.
\par
\begin{figure}[ht]
\centering
\includegraphics[width=0.8\columnwidth]{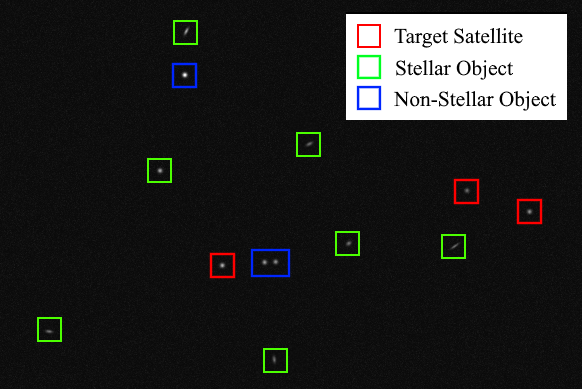}
\caption{A synthetic VBS image with point sources labelled.}
\label{exampleimage}
\end{figure}

When extending such frameworks to multiple targets, measurements at each epoch must consistently be assigned to corresponding targets if robust navigation is to be achieved. This requires the non-trivial ability to distinguish and identify multiple targets amongst all luminous spots in a VBS image (see Figure \ref{exampleimage}). Star identification algorithms can remove known stars from consideration, but there may also be stellar objects (SO) not in the on-board star catalog; non-stellar objects (NSO) such as other satellites or debris; or sensor defects such as hotspots. Relative state estimates can be used to identify targets but their existence or quality is not guaranteed. State initializations often possess significant uncertainty, meaning that several bearing angles in an image could be candidates for the target measurement. Furthermore, errors in measurement assignment compound errors in the state estimate, and vice versa. In the far range case ($\geq$1 km separation) considered in this paper, it is also impossible to use visual appearance for identification. The ability to independently, reliably track multiple objects across a sequence of images is therefore necessary to enable complete angles-only swarm navigation architectures. This must be achieved without requiring a-priori target state information if autonomy and self-initialization of navigation on-board is desired.
\par
Spaceflight also places stringent requirements on performance. Angles-only navigation exploits small nonlinearities to estimate spacecraft states \cite{josh} and is therefore particularly sensitive to measurement errors. A single incorrect assignment can cause divergence in the resulting state estimate, meaning close to 100\% assignment precision is necessary for risk-averse on-orbit applications. Furthermore, onboard resources are limited such that high computational efficiency is needed. Measurement frequencies are low, on the order of minutes between images, and use of visual measurements implies large data gaps when targets are in eclipse. Examples of applications under these constraints are the upcoming NASA Starling mission, which consists of four CubeSats in low Earth orbit (LEO) and intends to be the first flight demonstration of autonomous angles-only swarm navigation \cite{starling, artms}, and a proposed CubeSat swarm taking distributed measurements of the Martian atmosphere and ionosphere \cite{marsao}.
\par
More broadly, this field is referred to as multitarget tracking (MTT). In terrestrial contexts, a variety of MTT algorithms see frequent usage, including global nearest neighbour (GNN); joint probabilistic data association \linebreak (JPDA); multi-hypothesis tracking (MHT); and random finite set (RFS) methods, commonly in the form of a probability hypothesis density (PHD) filter \cite{mtt0, mtt1, mtt4, mtt2}. When considered for spaceborne angles-only tracking, each has particular advantages and disadvantages. GNN is simple but susceptible to poor performance when targets are not well-separated \cite{mtt1}. JPDA, though demonstrably accurate in a variety of scenarios, generally assumes a known number of targets \cite{mtt4}. Both approaches are then unideal in that targets may not be well-separated and the number of visible targets may be unknown. MHT is a theoretically optimal approach that performs well for low signal-to-noise ratios. However, it relies on forming increasing numbers of target track hypotheses such that heuristic hypothesis pruning is necessary for reasonable computation \cite{mtt1}. This is particularly challenging for low-powered spacecraft processors. RFS techniques are newer, with many promising varieties seeing continued development \cite{mtt0}; conversely, this makes them somewhat less proven and approximations are generally needed for real-time usage. Finally, although machine learning approaches to MTT have become increasingly popular, there are difficulties in generating visual training data representative of the space environment \cite{mtt3}.
\par
Relevant examples in hardware-limited terrestrial robotics are presented by Cano et al. \cite{mtt5}, who employ a Gaussian mixture PHD filter to track several robots with a single camera, and Farazi et al. \cite{mtt6}, who employ a neural-network-based pipeline running in real time on an observer robot. Both display promising accuracy, but rely on more detailed, higher-frequency imagery than is typical for spaceborne cameras \cite{starling}. A in-orbit example is given by LeGrand et al.  \cite{mtt7}, who suggest using a cardinalized PHD filter to track nearby NSO from an inspector spacecraft. However, they apply range information from stereo imagery which may not always be obtainable.
\par
For the case of a single target spacecraft, two prior flight experiments have conducted angles-only navigation. In 2012, the ARGON experiment \cite{argon} enabled the rendezvous of two smallsats in LEO from inter-satellite separations of 30km to 3km. To identify the target, bearing angles in successive images were linked by finding similarities in the sizes and positions of their associated pixel clusters. The target was assumed to move significantly less than other objects, and its measurement track was chosen as that displaying the largest difference from the average linked motion. In 2016, the AVANTI \cite{avanti} experiment performed a rendezvous of one smallsat and one picosat from separations of 13km to 50m. Unlike ARGON, which utilised ground-in-the-loop elements, AVANTI operated primarily autonomously. The Density-Based Spatial Clustering of Applications with Noise (DBSCAN) algorithm was applied for target identification. In sets of successive images, DBSCAN identified the target as a cluster of multiple measurements within a small radius, since it was again assumed to move less than other objects. Target tracks were expected to display curving motion, so measurements with outlying residuals after Bezier curve fitting were rejected. While ARGON and AVANTI proved successful, their measurement assignment methods assume a single target and cannot be applied to a multitarget scenario without significant modification.
\par
Bearing in mind the limitations of prior flight projects and current MTT methods, this paper develops an angles-only MTT algorithm suitable for spacecraft swarms. It fuses the inspirations of ARGON and AVANTI with a multi-hypothesis framework, in that the kinematics of relative orbits can be leveraged to enhance the robustness and reduce the complexity of a proven MHT approach. Henceforth, the algorithm is referred to as the `Spacecraft Angles-only MUltitarget tracking System' (SAMUS). SAMUS is agnostic to orbit eccentricity and requires only 1) coarse absolute orbit knowledge of the observer and 2) knowledge of the magnitudes and execution times of swarm maneuvers (but not which maneuvers correspond to which targets). No knowledge of the number of targets or their relative orbits is needed. In this fashion it improves upon existing algorithms to meet the tight requirements of general autonomous swarm operations in space.
\par
The following contributions are presented. First, a novel measurement transform reduces distortions in target tracks created by perturbing forces such as $J_2$ gravity. Target motion then has known form in the chosen observer reference frame. Second, knowledge of orbit dynamics is applied to fit parametric motion models to target tracks by solving pairs of linear systems. These models are used to predict target measurements in new images. Third, a set of kinematic rules is derived, used to describe target motion, reject unlikely measurement assignment hypotheses, and choose likely hypotheses. Fourth, an implementation of MHT is presented which aims to maintain a low computational footprint and high accuracy under challenging conditions. Fifth, tracking in the presence of known or partially-known swarm maneuvers is developed. Finally, SAMUS is validated with rigorous, high-fidelity test suites, comprising of Monte-Carlo simulations using both synthetic measurements and hardware-in-the-loop (HIL) imagery.
SAMUS can be integrated with state estimation frameworks to form a complete autonomous swarm navigation solution and will be flight tested in this form aboard the NASA Starling mission in 2022 \cite{starling, artms}.
\par
After this introduction, Section \ref{background} discusses the mathematical background of target relative orbit behavior and MHT. The detailed processes and reasoning behind the new algorithm are highlighted in Section \ref{algorithm}, followed by performance testing and validation in Section \ref{results}. Section \ref{conclusion} presents concluding remarks.

\section{Background}
\label{background}
\subsection{Coordinate Frames}
\noindent To obtain measurements, the observer spacecraft processes VBS images to compute the time-tagged bearing angles to objects in its field of view (FOV). Bearing angles consist of azimuth and elevation $(\alpha, \epsilon)^\top$ and subtend the line-of-sight (LOS) vector  $\delta \bm{r}^{\mathcal{V}} = (\delta r_x^{\mathcal{V}}, \delta r_y^{\mathcal{V}}, \delta r_z^{\mathcal{V}})^\top$ from the observer to its target. Superscript $\mathcal{V}$ indicates that the vector is described in the observer VBS coordinate frame. $\mathcal{V}$ consists of orthogonal basis vectors $\hat{\bm{x}}^\mathcal{V}, \hat{\bm{y}}^\mathcal{V}, \hat{\bm{z}}^\mathcal{V}$ where $\hat{\bm{z}}^\mathcal{V}$ is aligned with the camera boresight and $\hat{\bm{z}}^\mathcal{V} = \hat{\bm{x}}^\mathcal{V} \times \hat{\bm{y}}^\mathcal{V}$. This relates to bearing angles via \cite{josh}
\begin{equation}
\begin{bmatrix}
\alpha \\
\epsilon
\end{bmatrix}^{\mathcal{V}}
=
\begin{bmatrix}
\arcsin{(\delta r_y^{\mathcal{V}} / ||\delta \bm{r}^\mathcal{V}||_2)} \\
\arctan{(\delta r_x^{\mathcal{V}} / \delta r_z^{\mathcal{V}})}
\end{bmatrix}
\label{bearingangles}
\end{equation}
\noindent In this work, the primary observer reference frame is the observer radial/along-track/cross-track (RTN) frame $\mathcal{R}$. It is centered on and rotates with the observer and consists of orthogonal basis vectors $\hat{\bm{x}}^\mathcal{R}$ (directed along the observer's absolute position vector); $\hat{\bm{z}}^\mathcal{R}$ (directed along the observer's orbital angular momentum vector); and $\hat{\bm{y}}^\mathcal{R} = \hat{\bm{z}}^\mathcal{R} \times \hat{\bm{x}}^\mathcal{R}$ \cite{vallado}. Similarly, define a frame $\mathcal{W}$ using $\hat{\bm{y}}^\mathcal{W}$ (directed along the observer's velocity vector); $\hat{\bm{z}}^\mathcal{W} = \hat{\bm{z}}^\mathcal{R}$; and $\hat{\bm{x}}^\mathcal{W} = \hat{\bm{y}}^\mathcal{W} \times \hat{\bm{z}}^\mathcal{W}$. $\mathcal{W}$ only differs from $\mathcal{R}$ by a rotation of the observer flight path angle $\phi_f$ about $\hat{\bm{z}}^\mathcal{R}$, with $\phi_f \approx 0$ in near-circular orbits \cite{vallado}.
\par
\begin{figure}[ht]
\centering
\includegraphics[width=0.9\columnwidth]{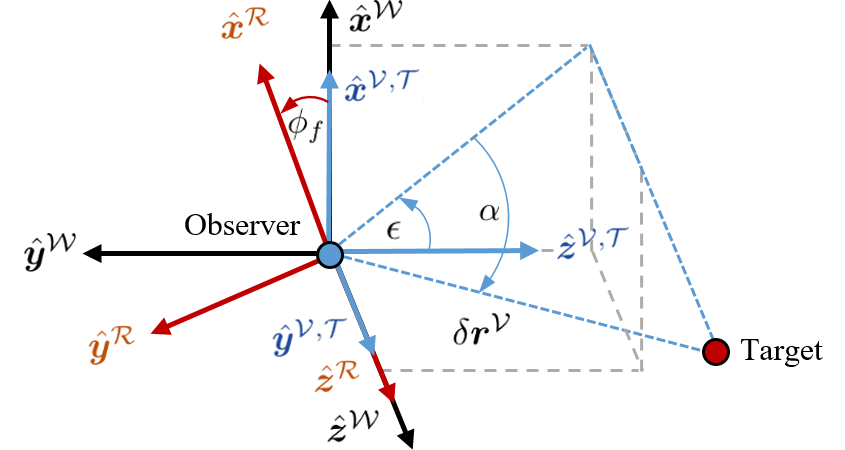}
\caption{Definition of target LOS vector and bearing angles with respect to observer coordinate frames $\mathcal{V}, \mathcal{T}, \mathcal{R}$ and $\mathcal{W}$.}
\label{frames}
\end{figure}

Typical angles-only navigation scenarios present targets with large separations in the velocity or anti-velocity directions \cite{marsao, starling, argon, avanti}. Thus, when defining the tracking frame $\mathcal{T}$ in which MTT is performed, a natural choice is to align its basis vector $\hat{\bm{z}}^\mathcal{T}$ with the observer's velocity or anti-velocity direction $\pm \hat{\bm{y}}^\mathcal{W}$. Consequently, $\hat{\bm{y}}^\mathcal{T}$ is aligned with the observer's orbital angular momentum vector and $\hat{\bm{x}}^\mathcal{T} = \hat{\bm{y}}^\mathcal{T} \times \hat{\bm{z}}^\mathcal{T}$. $\mathcal{T}$ then differs from $\mathcal{W}$ by a rotation of +90\degree about $\hat{\bm{x}}^\mathcal{W}$. For convenience, we align the VBS frame $\mathcal{V}$ with $\mathcal{T}$; otherwise, LOS vectors in $\mathcal{V}$ can be rotated into $\mathcal{T}$ using rotation matrices with respect to the Planet-Centered Inertial (PCI) frame $\mathcal{P}$, as per
\begin{equation}
\delta \bm{r}^{\mathcal{T}} =\ ^\mathcal{W} \overrightarrow{\bm{R}}^\mathcal{T} \ ^\mathcal{P} \overrightarrow{\bm{R}}^\mathcal{W} \ ^\mathcal{V}\overrightarrow{\bm{R}}^\mathcal{P} \ \delta \bm{r}^{\mathcal{V}}
\end{equation}
\noindent where $^\mathcal{A} \overrightarrow{\bm{R}}^\mathcal{B}$ denotes a rotation from frame $\mathcal{A}$ into frame $\mathcal{B}$. $^\mathcal{W} \overrightarrow{\bm{R}}^\mathcal{T} $ is known from geometry;  $^\mathcal{P} \overrightarrow{\bm{R}}^\mathcal{W}$ is known if the observer's absolute orbit is being estimated; and $^\mathcal{V}\overrightarrow{\bm{R}}^\mathcal{P}$ is computed by performing star identification and attitude determination with the VBS \cite{argon}. Figure \ref{frames} depicts the relationships between frames and measurements.
\par
To capture orbit curvature with improved accuracy, the relative positions of targets with respect to an observer can be described in curvilinear coordinates $\delta \bm{r}_{\textrm{curv}}^{\mathcal{R}} = (\delta r, a\Theta, a\Phi)$. Here, $\delta r, \Theta, \Phi$ are differences in orbit radii, angular in-plane separations and angular out-of-plane separations respectively, for observer semimajor axis $a$ \cite{maneuverfree}. Hereafter the curvilinear representation is used, which can be mapped back to rectilinear coordinates via
\begin{equation}
\delta \bm{r}_{\textrm{rect}}^{\mathcal{R}}
=
\begin{bmatrix}
(a + \delta r) c_{\Theta} c_{\Phi} - a \\
(a + \delta r) s_{\Theta} c_{\Phi} \\
(a + \delta r) s_{\Phi}
\end{bmatrix}
\label{mapping}
\end{equation}
\noindent $c$ and $s$ denote cosine and sine of the subscripted argument.
\subsection{Relative Orbit Dynamics}
\label{dynamicsbg}
\noindent To discuss the dynamic behaviour of targets, this paper applies a quasi-nonsingular relative orbital element (ROE) state representation. The ROE are defined in terms of the absolute orbital elements (OE) of the observer and target (denoted by subscripts `$o$' and `$t$' respectively) as \cite{damicothesis}
\begin{equation}
\medmuskip=0mu
\thinmuskip=0mu
\thickmuskip=1mu
\delta \bm{x}_{\textrm{roe}} = 
\begin{bmatrix}
\delta a \\
\delta \lambda \\
\delta e_x \\
\delta e_y \\
\delta i_x \\
\delta i_y
\end{bmatrix}
=
\begin{bmatrix}
\delta a \\
\delta \lambda \\
|\delta \bm{e}| \, c_{\phi} \\
|\delta \bm{e}| \, s_{\phi} \\
|\delta \bm{i}| \, c_{\theta} \\
|\delta \bm{i}| \, s_{\theta}
\end{bmatrix}
=
\medmuskip=0mu
\thinmuskip=0mu
\thickmuskip=1mu
\begin{bmatrix}
(a_t - a_o) / a_o \\
(u_t - u_o) + c_{i_o}(\Omega_t - \Omega_o) \\
e_t c_{\omega_t} - e_o c_{\omega_o} \\
e_t s_{\omega_t} - e_o s_{\omega_o} \\
i_t - i_o \\
s_{i_o}(\Omega_t - \Omega_o)
\end{bmatrix}
\label{ROEs}
\end{equation}
\noindent Above, $a, e, i, \Omega, \omega$ and $M$ are the classical Keplerian OE and $u = M + \omega$ is the mean argument of latitude. The ROE consist of $\delta a$, the relative semimajor axis; $\delta \lambda$, the relative mean longitude (analogous to target range); $\delta \bm{e}$, the relative eccentricity vector with magnitude $\delta e$ and phase $\phi$; and $\delta \bm{i}$, the relative inclination vector with magnitude $\delta i$ and phase $\theta$. This representation is singular for equatorial orbits and fully nonsingular ROE have also been developed \cite{nonsingular}. Furthermore, for improved application to eccentric orbits, Sullivan et. al. \cite{josh} present `eccentric ROE' (EROE). The EROE feature a modified relative mean longitude $\delta \lambda^*$ and modified relative eccentricity vector $\delta \bm{e}^*$ (with magnitude $\delta e^*$ and phase $\phi^*$), defined by
\begin{align}
\delta \lambda^* &= \xi \delta \lambda + (1 - \xi) \Big(\mkern-6mu - \frac{s_{\omega_o}}{e_o} \delta e_x + \frac{c_{\omega_o}}{e_o} \delta e_y + \cot{i_o \delta i_y}\Big)
\label{EROE1}
\\
\delta e_x^* &= \frac{c_{\omega_o} \delta e_x + s_{\omega_o} \delta e_y}{1 - e_o^2} = \delta e^* c_{\phi^*}
\label{EROE2}
\\
\delta e_y^* &= \frac{e_o (-\delta \lambda \mkern-2mu + \mkern-2mu \cot{i_o \delta i_y}) \mkern-2mu - \mkern-2mus_{\omega_o} \delta e_x \mkern-2mu+\mkern-2mu c_{\omega_o} \delta e_y}{(1 - e_o^2)^{3/2}} \mkern-2mu=\mkern-2mu \delta e^* s_{\phi^*}
\label{EROE3}
\\
\xi &= \frac{(1 + e_o^2 / 2)}{(1 - e_o^2)^{3/2}}
\end{align}
\noindent The EROE reduce to the ROE for $e = 0$. %
\par
\begin{figure}[ht]
\centering
\includegraphics[width=0.9\columnwidth]{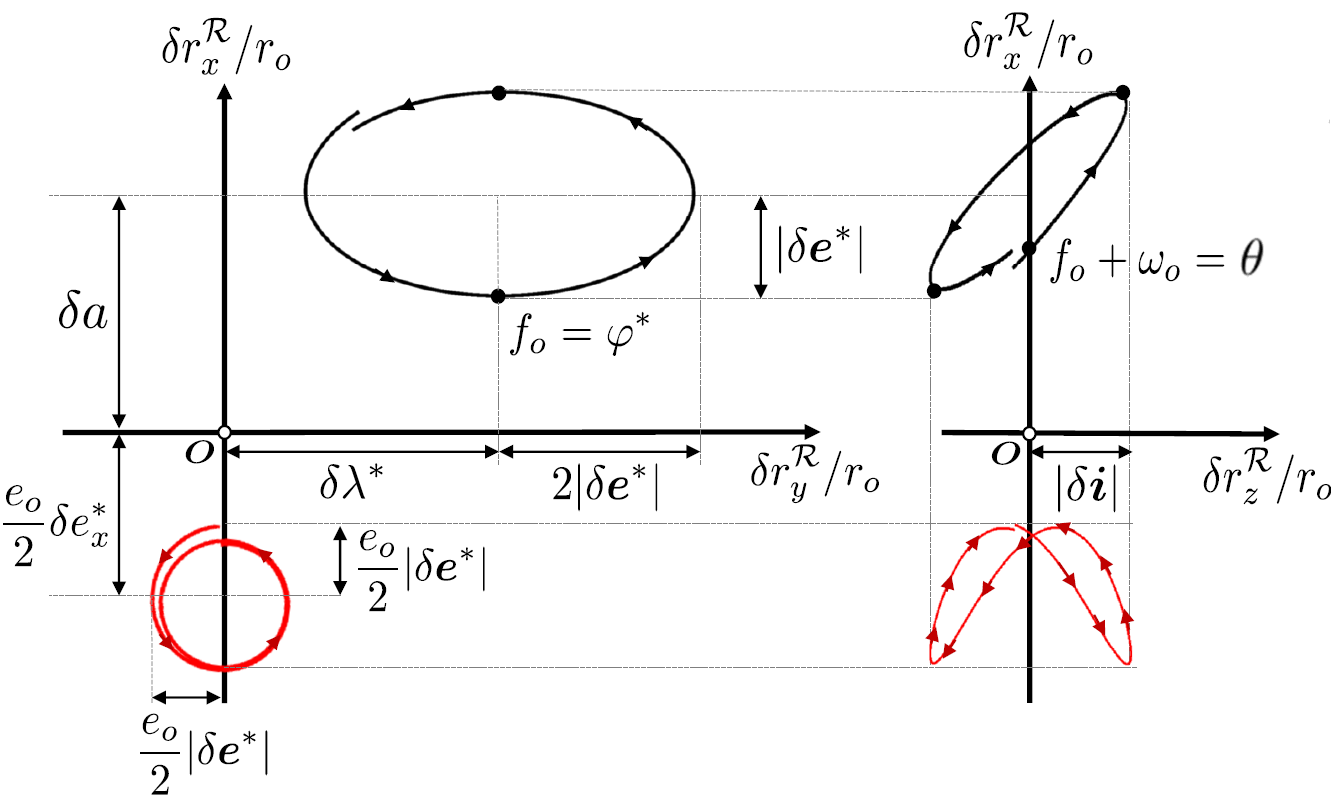}
\caption{Target relative motion in the $\hat{\bm{x}}^{\mathcal{R}}$-$\hat{\bm{y}}^{\mathcal{R}}$ (RT) and $\hat{\bm{x}}^{\mathcal{R}}$-$\hat{\bm{z}}^{\mathcal{R}}$ (RN) planes \cite{generalized}. Motion that is first-order in spacecraft separation is in black. Contributions proportional to $e_o$ are in red.}
\label{ellipse}
\end{figure}

A particularly useful aspect of the ROE are that they provide geometric intuition regarding target relative motion. This was first demonstrated by D'Amico for near-circular orbits \cite{damicothesis}, who formulated a linear map between the ROE and the target's nondimensional RTN relative position. 
Subsequently, Sullivan et al. \cite{josh} mapped the EROE to target RTN position as per
\begin{equation}
\renewcommand{\arraystretch}{1.2}
\begin{bmatrix}
 \delta r_x^{\mathcal{R}}\\
 \delta r_y^{\mathcal{R}}\\
 \delta r_z^{\mathcal{R}}
\end{bmatrix}
\medmuskip=1mu
\thinmuskip=1mu
\thickmuskip=3mu
\approx
r_o
\begin{bmatrix}
\delta a - \frac{e_o}{2} \delta e_x^* - \delta e^* \Big(c_{f_o - \phi^*} + \frac{e_o}{2} c_{2f_o - \phi^*}\Big)\\
\delta \lambda^* + \delta e^* \Big(2 s_{f_o - \phi^*} + \frac{e}{2} s_{2f_o - \phi^*} \Big)\\
\delta i s_{f_o + \omega_o - \theta}
\end{bmatrix}
\label{VBSmotion}
\end{equation}
\noindent 
Figure \ref{ellipse} presents relative motion in RTN for small separations \cite{generalized}. Oscillatory motion produced by target relative orbits is shown in black, possessing the same frequency as the orbit. Oscillatory motion produced by orbit eccentricity is shown in red, acting at twice the frequency of the orbit. Thus, $\delta a$ and $\delta \lambda^*$ capture mean offsets in the radial and along-track directions respectively; magnitudes of $\delta e^*$ and $\delta i$ correspond to magnitudes of oscillations in the RT and RN planes respectively; and phases of $\delta e^*$ and $\delta i$ dictate the orientation and aspect ratio of the tilted ellipse in the RN plane. The eccentricity of the observer's orbit superimposes additional offsets and higher-frequency oscillations in the RT and RN planes.
\par
For angles-only tracking, the relevant components of motion are those occurring in the image plane of the VBS. As per the earlier coordinate frame definitions, these are components $(\delta r_x^{\mathcal{R}}, \delta r_z^{\mathcal{R}})$ in Equation \ref{VBSmotion} and Figure \ref{ellipse}. The elliptical aspects of this motion can be described using the traditional geometric ellipse parameters of semimajor axis $a_e$, semiminor axis $b_e$, center $(x_e, y_e)$ and tilt $\gamma_e$ via \cite{maneuverfree}
\begin{align}
(x_e, y_e) &= (\delta a, 0)\\
(a_e, b_e) &= \bigg(\frac{\delta e^2 \mkern-6mu + \mkern-2mu \delta i^2 \mkern-6mu \pm \mkern-3mu \sqrt{\delta e^4 \mkern-6mu + \mkern-2mu \delta i^4 \mkern-6mu - \mkern-2mu 2\delta e^2 \delta i^2 c_{2(\phi - \theta)}}}{2}\bigg)^{\frac{1}{2}}\\
\gamma_e &= \frac{1}{2}\arctan{\Big(\frac{-2 \delta e \delta i s_{\phi - \theta}}{\delta e^2 - \delta i^2} \Big)}
\label{ellipse3}
\end{align}
\par
Target relative motion is also affected by disturbing forces such as atmospheric drag, solar radiation pressure (SRP), third-body gravity and spherical harmonic gravity. These cause secular drifts, long-period perturbations and short-period perturbations to target motion \cite{damicothesis}. On the timescales of image-to-image tracking (i.e. $\leq 5$ minutes) short-period perturbations are particularly detrimental. In LEO, the most significant perturbation is generally $J_2$ Earth oblateness, whose effects are commonly orders of magnitude larger than other disturbances \cite{damicothesis}. Its short-period and secular effects on the ROE are \cite{linearised} 
\renewcommand{\arraystretch}{1.2}
\begin{align}
\delta \bm{e}_{\textrm{sp}} \mkern-4mu &= \mkern-4mu
\begin{bmatrix}
\delta e_{x, \textrm{sp}}\\
\delta e_{y, \textrm{sp}}
\end{bmatrix}
\mkern-4mu = \mkern-4mu
\frac{3 J_2 R_P^2}{2 a^2} \mkern-3mu
\begin{bmatrix}
(1 \mkern-3mu - \mkern-3mu \frac{5}{4} s_i^2) c_{u} \mkern-3mu + \mkern-3mu (\frac{7}{12} s_i^2) c_{3u}  \\
 (1 \mkern-3mu - \mkern-3mu \frac{7}{4} s_i^2) s_{u} \mkern-3mu + \mkern-3mu (\frac{7}{12} s_i^2) s_{3u} 
\end{bmatrix}
\label{spdistort} \\
\delta\bm{i}_{\textrm{sp}} \mkern-4mu &= \mkern-4mu
\begin{bmatrix}
\delta i_{x, \textrm{sp}}\\
\delta i_{y, \textrm{sp}}
\end{bmatrix}
\mkern-4mu = \mkern-4mu
\frac{3 J_2 R_P^2}{8 a^2}
\begin{bmatrix}
s_{2i} \, c_{2u} \\
2 c_{i} \, s_{i} \, s_{2u}
\end{bmatrix}
\label{spdistort1}
\end{align}
\renewcommand{\arraystretch}{1.2}
\begin{align}
\delta \bm{e}_{\textrm{sec}} \mkern-4mu &= \mkern-4mu
\begin{bmatrix}
\delta e_{x, \textrm{sec}}\\
\delta e_{y, \textrm{sec}}
\end{bmatrix}
\mkern-4mu = \mkern-2mu
\delta e 
\begin{bmatrix}
\cos{(\phi_0 + \frac{3\pi t}{2T} J_2 \frac{R_P^2}{a^2} (5 c_i^2 \mkern-4mu - \mkern-4mu 1))} \\
\sin{(\phi_0 +  \frac{3\pi t}{2T} J_2 \frac{R_P^2}{a^2} (5 c_i^2 \mkern-4mu - \mkern-4mu 1))}  
\end{bmatrix}
\label{secdistort1}
\\
\delta \bm{i}_{\textrm{sec}} \mkern-4mu &= \mkern-4mu
\begin{bmatrix}
\delta i_{x, \textrm{sec}}\\
\delta i_{y, \textrm{sec}}
\end{bmatrix}
\mkern-4mu = \mkern-2mu
\delta i 
\begin{bmatrix}
1 \\
1 \mkern-4mu - \mkern-4mu \frac{3 \pi t}{T} J_2 \frac{R_E^2}{a^2} \delta i s_i^2
\end{bmatrix}
\label{secdistort}
\end{align}
\noindent where $t$ is time, $T$ is the orbit period, $R_P$ is the radius of the central body and $\phi_0$ is the phase of $\delta \bm{e}$ at initial epoch $t_0$. These variations in $\delta \bm{e}$ and $\delta \bm{i}$ must be included in Equation \ref{VBSmotion} if their effects are significant in the orbit regime of interest. Similar expressions have been derived for other forces such as drag \cite{nonsingular}.
\par
Finally, maneuvers by the observer or targets also affect relative motion in $\mathcal{R}$. Consider a change in the ROE, $\Delta \delta \bm{x}_{\textrm{roe}}$, and a maneuver by the observer in RTN, $\delta \bm{v}_o^{\mathcal{R}} = (\delta v_x^{\mathcal{R}}, \delta v_y^{\mathcal{R}}, \delta v_z^{\mathcal{R}})^\top$. In eccentric orbits, these are related via a control input matrix $\bm{B}_{\textrm{roe}}$ \cite{josh, michelle}, defined as
\begin{align}
&\Delta \delta \bm{x}_{\textrm{roe}} = \bm{B}_{\textrm{roe}} \delta \bm{v}_o^{\mathcal{R}}\\
\renewcommand{\arraystretch}{1.4}
&\bm{B}_{\textrm{roe}} =
-\frac{\eta}{a n}
    \setlength\arraycolsep{2pt}
    \begin{bmatrix}
    \frac{2 e}{\eta^2} s_{f} & \frac{2k}{\eta^2} & 0\\
    \frac{(\eta - 1) k c_{f} - 2\eta e}{e k} & \frac{(1 - \eta)(k + 1) s_{f}}{e k} & 0\\
    s_{f + \omega} & \frac{(k + 1)c_{f + \omega} + e_x}{k} & \frac{e_y s_{f + \omega}}{k \tan{i}}\\
    -c_{f + \omega} & \frac{(k + 1)s_{f + \omega} + e_y}{k} & -\frac{e_x s_{f + \omega}}{k \tan{i}}\\
    0 & 0 & \frac{c_{f + \omega}}{k}\\
    0 & 0 & \frac{s_{f + \omega}}{k}
    \end{bmatrix}
    \label{control}\\
&\eta = \sqrt{1 - e^2}, \quad n = \sqrt{\mu / a^3}, \quad k = 1 + e c_f
\end{align}
\noindent Similarly, change in ROE from a target maneuver is obtained via $\Delta \delta \bm{x}_{\textrm{roe}} = -\bm{B}_{\textrm{roe}} \delta \bm{v}_t^{\mathcal{R}}$. Change in the EROE, $\Delta \delta \bm{x}_{\textrm{eroe}}$, is computed by first computing $\Delta \delta \bm{x}_{\textrm{roe}}$ and then mapping this to $\Delta \delta \bm{x}_{\textrm{eroe}}$ via Equations \ref{EROE1} - \ref{EROE3}.
\subsection{Multi-Hypothesis Tracking}
\label{mhtbg}
\begin{figure}[ht]
\centering
\includegraphics[width=0.9\columnwidth]{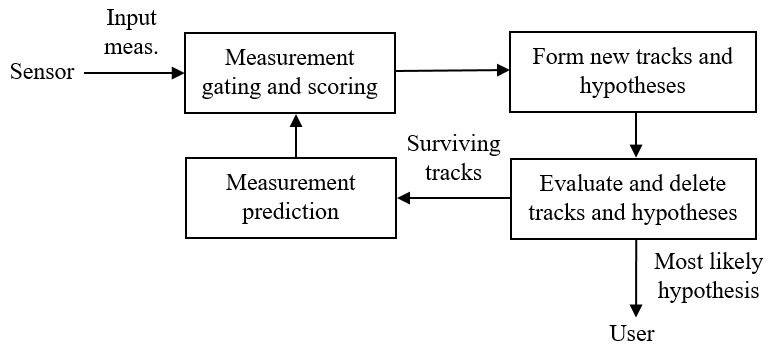}
\caption{MHT logic overview \cite{mht0}.}
\label{logic}
\end{figure}

\noindent The central objective of MTT is to collect sensor data containing one or more potential targets and to partition it into sets of observations $-$ or `tracks' $-$ produced by the same target over time \cite{mht0}. Assume that tracks have been formed from previous data and that a new set of measurements $-$ or `scan' $-$ has become available. Then, a typical MTT system performs five sequential tasks:
\begin{enumerate}[itemsep=-1mm]
\item Sensor data processing: retrieve new measurements.
\item Measurement prediction: use existing tracks to predict new measurements.
\item Measurement gating: assess which new measurements may reasonably be assigned to which tracks.
\item Measurement-to-track association: score valid assignments and determine the best option(s).
\item Track maintenance: initialize new tracks, confirm likely tracks and delete unlikely tracks.
\end{enumerate}
\noindent Difficulties arise when targets are closely spaced and multiple observations may be assigned to multiple tracks $-$ the correct choice can be challenging to determine. A prominent approach, leveraged by SAMUS, is multi-hypothesis tracking. MHT applies a delayed decision philosophy by propagating and maintaining multiple assignment hypotheses, since future data can aid in disambiguating past assignments. The operational logic is presented in Figure \ref{logic}. With each new scan, new measurements are received and are gated with respect to existing tracks. New tracks and hypotheses are then formed and evaluated in terms of likelihood. Finally, unlikely hypotheses are deleted and new measurements are predicted for surviving tracks. MHT was initially developed by Reid \cite{mht1} and has since been expanded into a variety of forms \cite{mht2, mht3, mht4, mht5a}.
\par
\begin{figure}[ht]
\centering
\includegraphics[width=0.9\columnwidth]{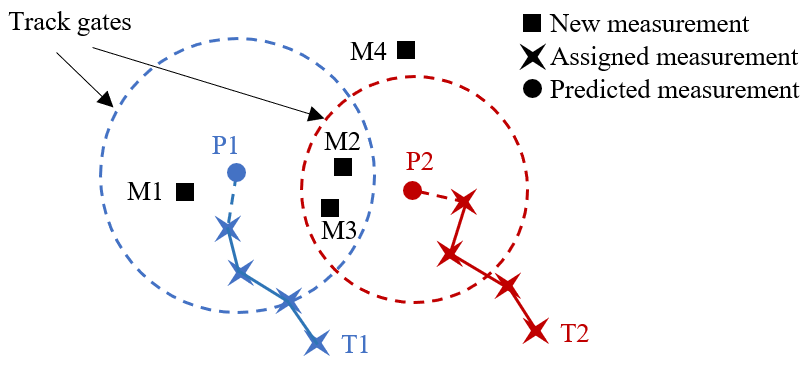}
\caption{An example of measurement assignment ambiguity. Measurements M2 and M3 lie within the track gates of both targets.}
\label{observations}
\end{figure}
Consider Figure \ref{observations}, in which tracks T1 and T2 lead to predicted observations P1 and P2. Four measurements are received: M1, ..., M4. Measurements can be associated with existing tracks if they fall within track gates, or alternatively can start a new track. Following the convention of Blackman \cite{mht0}, denote T3 (T1, M1) as Track 3 formed from the association of T1 and M1. Similarly, there exists T4 (T1, M2); T5 (T1, M3); T6 (T2, M2); and T7 (T2, M3). Furthermore, NT1, ..., NT4 denote new tracks initiated from M1, ..., M4. Tracks are `compatible' if they have no observations in common, and MHT `hypotheses' are composed of sets of compatible tracks. In the above example there are 10 feasible hypotheses, including H1: (T1, T2, NT1, ..., NT4), H2: (T3, T6, NT3, NT4), H3: (T3, T7, NT2, NT4), and so on. Upon receiving new measurements, existing hypotheses are expanded into sets of new hypotheses by considering all valid measurement-to-track assignments that maintain compatibility.
\par
To evaluate hypotheses, MHT must account for physical consistency as well as probability of target presence or false alarms. The likelihood ratio (LR) for collecting data into a track is traditionally defined as \cite{mht6, mht7}
\begin{align}
\textrm{LR} = \frac{p(D|H_1)P_0(H_1)}{p(D|H_0)P_0(H_0)} = \frac{P_T}{P_F}
\label{MHTscore}
\end{align}
\noindent Hypotheses $H_1$ and $H_0$ are the true target and false alarm hypotheses with probabilities $P_T$ and $P_F$ respectively. $P_T$ assumes all track observations are of the same target, and $P_F$ assumes all track observations are of the background. $D$ represents data such that $p(D|H_i)$ is the probability density function evaluated with received data $D$ under the assumption that $H_1$ is correct. $P_0(H_i)$ is the a-priori probability of $H_i$. In practice the log-likelihood ratio (LLR) is generally used because it directly relates to true target probability $P_T$ via
\begin{equation}
\textrm{LLR} = \ln{(P_T | P_F)} \implies P_T = e^{\textrm{LLR}} / (1 + e^{\textrm{LLR}})
\end{equation}
\noindent The LLR is also known as the track score, and the score of a hypothesis is the sum of all constituent track scores. To present an output to the user, MHT can simply provide the most likely track per target. This, however, can lead to inconsistencies in the output hypothesis as track probabilities change with the receipt of more data. Alternately, it may provide an average state estimate and covariance computed from all branch probabilities, but this does not always correspond to an actual set of measurements.
\par
\begin{figure}[ht]
\centering
\includegraphics[width=0.9\columnwidth]{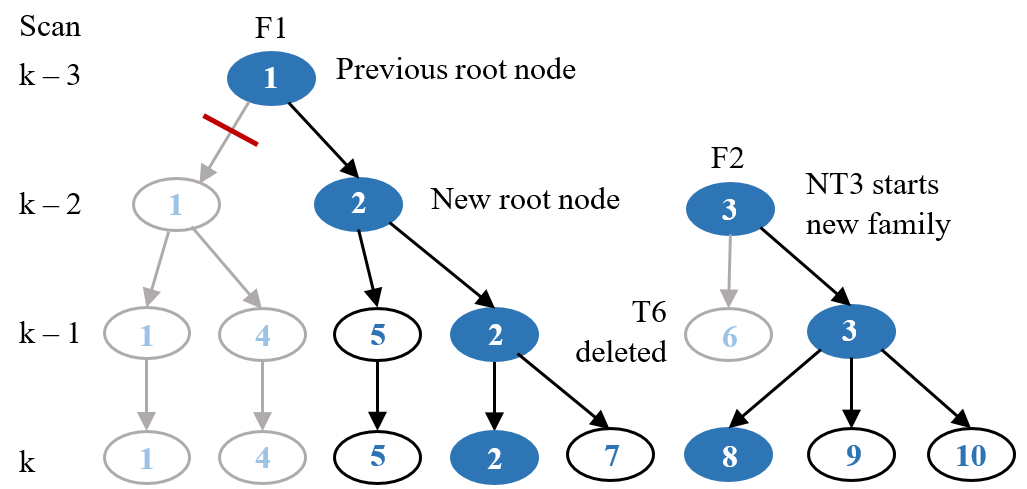}
\caption{Target trees with root node updates \cite {mht0, mht2}. Blue denotes the best global hypothesis. The new root node is selected at $k - 2$ and dissimilar branches are pruned. T6 was previously deleted for being a low-scoring track.}
\label{tree}
\end{figure}

A prominent disadvantage of MHT is the potential combinatorial explosion in the number of generated tracks and hypotheses as new scans arrive \cite{mht0}. Track pruning and merging must therefore be used to control growth. When describing these operations, tracks may be viewed as branches in a tree: nodes occur when tracks split into multiple hypotheses, and a `tree' is a set of tracks with a common root node that represents one hypothesized target. `Root node updating' or `N-scan pruning' determines which tracks in each tree are part of the best current hypothesis (at step $k$) and goes back $N$ scans (e.g. $N = 2$) to establish a new root node. Figure \ref{tree} presents a new example with trees F1 and F2. In F1, Track 2 is part of the best hypothesis and is established as the new root. Subsequently, hypotheses on the left-hand branch are discarded and decisions prior to scan $k - 2$ are considered final. New global hypotheses are then formed by choosing at most one track from each tree: for example, H1 (T2, T10).
\par
Clustering, the $m$-best method and track-oriented MHT are also used to reduce the number of hypotheses for computation  \cite{mht0}. One `cluster' is a collection of all tracks which can be linked through common observations. Clusters can be processed independently since different clusters do not share any measurements or influences. By decomposing MHT into a set of smaller problems, it can be treated with fewer computations. The $m$-best algorithm applies Murty's method \cite{mht8} for finding the $m$-best solutions to the assignment problem. It limits the number of new hypotheses formed at scan $k$ to $m(k)$, preventing creation of many low-probability hypotheses. In track-oriented MHT, rather than maintaining and expanding hypotheses from scan-to-scan, existing hypotheses are discarded and new hypotheses are formed from tracks that survived pruning. This improves performance when there are many more hypotheses than tracks.
\section{SAMUS Algorithm}
\label{algorithm}
\noindent The angles-only MTT algorithm developed in this paper fuses the single-target kinematic techniques of prior flight projects with a multitarget MHT framework. 
SAMUS applies the core concept of MHT in that as measurements arrive, multiple tracks and hypotheses are simultaneously initialized, propagated, scored and trimmed, with the intention of robustly converging to the correct hypothesis over time. Novelty arises from the application of domain-specific knowledge to greatly improve the accuracy and efficiency of the approach. MHT is chosen as a basis
because it is considered mature and demonstrably accurate \cite{mht0, mht2} with its most significant disadvantage being the need to heuristically trim hypotheses for real-time computation. However, the generally consistent behavior of targets in orbit provides particularly effective trimming criteria and is well-suited to the `delayed decision' approach of MHT $-$ as more information is received, target motion can be judged more conclusively to arrive at the correct assignments. MHT is also able to quickly converge to a physical hypothesis, which is not always possible with the probabilistic estimates provided by other methods.
\par
\begin{figure}[ht]
\centering
\includegraphics[width=0.9\columnwidth]{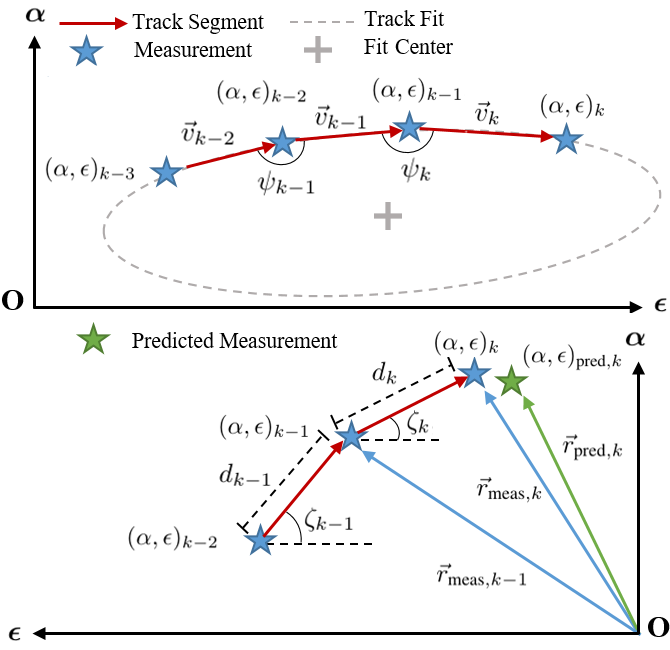}
\caption{Geometric target track quantities in bearing angle space, with elevation on the x-axis and azimuth on the y-axis. $(\alpha, \epsilon)_k$ are bearing angle measurements at epoch $k$.}
\label{baspace}
\end{figure}

Figure \ref{baspace} defines geometric quantities for SAMUS in 2D bearing angle space. In the upper figure, $(\alpha, \epsilon)_k$ is a track measurement at epoch $k$; the vector $\vec{v}_k$ is the `track step' from $(\alpha, \epsilon)_{k-1}$ to $(\alpha, \epsilon)_k$; and $\psi_k$ is the angle between $\vec{v}_{k-1}$ and $\vec{v}_k$. In the lower figure, $d_k$ is the magnitude of $\vec{v}_k$ while $\zeta_k$ is its phase; $\vec{r}_{\textrm{meas}, k}$ is the vector from the origin to $(\alpha, \epsilon)_k$; and $\vec{r}_{\textrm{pred}, k}$ is the vector from the origin to predicted measurement $(\alpha, \epsilon)_{\textrm{pred},k}$.
\par
\begin{figure}[ht]
\centering
\includegraphics[width=0.85\columnwidth]{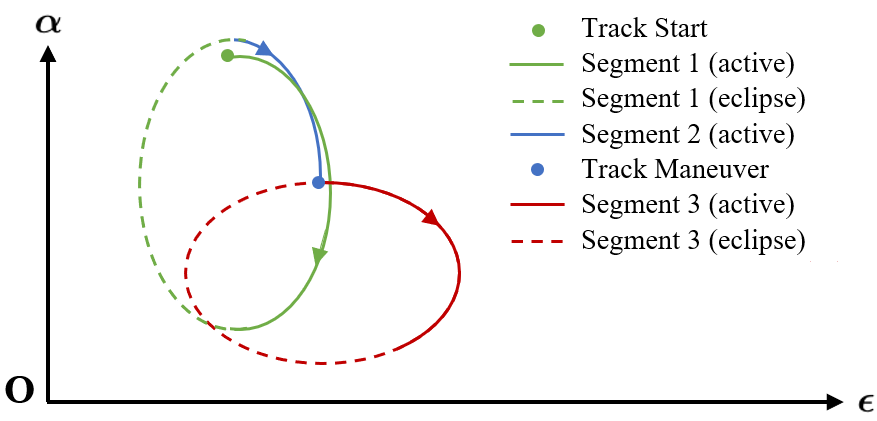}
\caption{Notional illustration of track segments.}
\label{segments}
\end{figure}

Within SAMUS, target tracks are separated into active `segments'. Segments are separated by either 1) a hypothesized maneuver, or 2) a measurement gap such as an eclipse period. When tracking within a segment, SAMUS applies knowledge that target motion should be consistent as per Equation \ref{VBSmotion}. Upon encountering a maneuver or measurement gap, SAMUS leverages the expected changes from Equations \ref{spdistort} - \ref{control} to assign the next measurement. Tracking then proceeds as normal during the subsequent segment. Figure \ref{segments} presents a notional illustration.
\par
To operate, three assumptions are required by SAMUS. First, targets remain sufficiently within the VBS FOV such that consistent measurement arcs of $\geq$4 successive images can be obtained per orbit. Second, the observer's absolute orbit is coarsely known such that the rotations between inertial and observer reference frames can be estimated and expected target visibility can be computed (e.g. expected orbit eclipse periods, or periods when the target may be out of the FOV). Third, maneuvers by the observer and targets during the tracking period are impulsive, and their execution times and RTN components are known. However, the correspondence between each specific maneuver and tracked target does not need to be known. This scenario is representative of a swarm mission with active control. The following sections present SAMUS in detail echoing the MTT task order from Section \ref{mhtbg}: 1) sensor data processing, 2) measurement prediction, 3) track gating, 4) measurement association, 5) maneuver association and 6) track maintenance.
\subsection{Sensor Data Processing}
\label{algorithm1}
\begin{figure*}[ht]
\centering
\includegraphics[width=6in, trim = {2cm 0cm 2cm 0cm}]{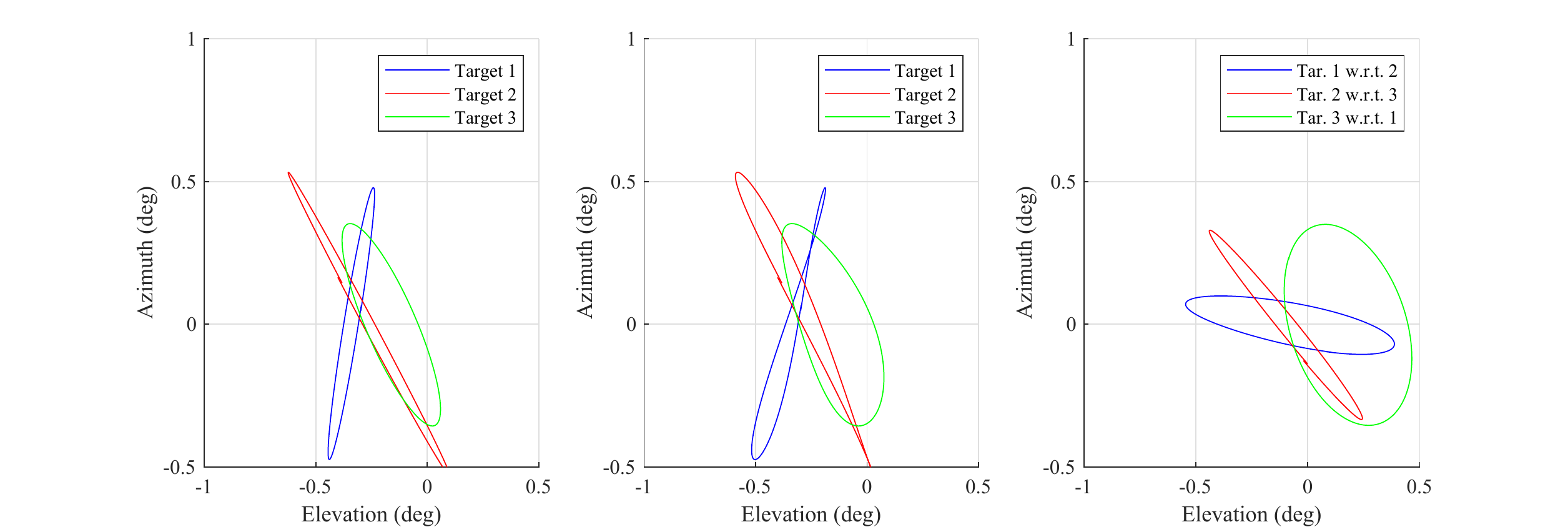}
\caption{Target trajectories without $J_2$ effects (left), with $J_2$ effects (center), and with $J_2$ effects and the measurement transform (right).}
\label{distortion}
\end{figure*}
\noindent A VBS image typically contains many luminous objects, including targets, other NSO, and SO. First, centroiding algorithms are used to simplify the raw image into a list of pixel cluster centroids. Star identification algorithms, such as the Pyramid algorithm \cite{pyramid} for lost-in-space scenarios, are applied to remove SO from the list of centroids. Non-catalog SO are detected by considering objects with unit vectors in the PCI frame which remain unchanged between images. Camera hotspots are removed by considering objects with constant pixel coordinates between images. These steps are common in star tracker usage \cite{argon, avanti} and are not detailed here, but result in a list of bearing angles to targets and other remaining unidentified objects in the FOV. This is the input MTT scan.
\par
Non-Keplerian forces (such as $J_2$ gravity in Equation \ref{spdistort}) affect target motion and distort the parametric form of Equation \ref{VBSmotion}. To reduce these effects, SAMUS uniquely applies a measurement transform when multiple targets are present. Tracks of a target $i$ are synchronously differenced with respect to tracks of a target $j$ ($i \neq j$), thus using $j$ as the virtual, moving origin of a tracking frame for $i$, as per
\begin{equation}
(\alpha, \epsilon)_i^{\mathcal{T}_{i/j}} = (\alpha, \epsilon)_i^{\mathcal{T}} - (\alpha, \epsilon)_j^{\mathcal{T}}
\end{equation}
\noindent Frame $\mathcal{T}_{i/j}$ denotes target $i$ viewed with respect to target $j$. In angles-only scenarios of interest, swarm members are relatively close together in inertial space and are affected similarly by perturbations. Thus, the origin of $\mathcal{T}_{i/j}$ and target $i$ measurements are affected by similar forces. In $\mathcal{T}_{i/j}$, perturbation effects are approximately cancelled and motion with the form of Equation \ref{VBSmotion} is recovered. More formally, consider targets $i$ and $j$ with similar OE such that short-period oscillations $\delta \bm{e}_{i, \textrm{sp}} \approx \delta \bm{e}_{j, \textrm{sp}}$ and $\delta \bm{i}_{i, \textrm{sp}} \approx \delta \bm{i}_{j, \textrm{sp}}$. Then, their $\hat{\bm{z}}^\mathcal{R}$-components of relative position are
\begin{align}
\delta r_{z, i}^\mathcal{R} &= (\delta i_{x_i} \mkern-3mu + \mkern-2mu \delta i_{x_i, \textrm{sp}}) s_{f_o + \omega_o} \mkern-2mu - (\delta i_{y_i} \mkern-3mu + \mkern-2mu \delta i_{y_i, \textrm{sp}}) c_{f_o + \omega_o}\\
\delta r_{z, j}^\mathcal{R} &= (\delta i_{x_j} \mkern-3mu + \mkern-2mu \delta i_{x_j, \textrm{sp}}) s_{f_o + \omega_o} \mkern-2mu - (\delta i_{y_j} \mkern-3mu + \mkern-2mu \delta i_{y_j, \textrm{sp}}) c_{f_o + \omega_o}\\
\implies &\delta r_{z, i}^{\mathcal{R}_{i/j}} = \delta r_{z, i}^\mathcal{R} \mkern-3mu - \mkern-2mu \delta r_{z, j}^\mathcal{R} \\
& \approx (\delta i_{x_i} \mkern-3mu - \mkern-2mu \delta i_{x_j}) s_{f_o + \omega_o} \mkern-3mu - \mkern-2mu (\delta i_{y_i} \mkern-3mu - \mkern-2mu \delta i_{y_j}) c_{f_o + \omega_o}
\end{align}
\noindent and $\delta r_{z, i}^{\mathcal{R}_{i/j}}$ recovers the form of Equation \ref{VBSmotion}. Similar results are obtained for the $\hat{\bm{x}}^\mathcal{R}$ and  $\hat{\bm{y}}^\mathcal{R}$ components of relative motion. Effects of this transform for the case in Table \ref{distortROEs} are shown in Figure \ref{distortion}.
\begin{table}[ht]
\renewcommand{\arraystretch}{1.0}
\caption{Swarm configuration for Figure \ref{distortion}.}
\label{distortROEs}
\centering
\begin{tabular}{|c|c||c|c|c|c| }
\hline
OE & Obser. & ROE & Tar. 1 & Tar. 2 & Tar. 3 \\
\hline
$a$ (km) 				& 6878 & $\delta a$ (km) 		& -0.1 & -0.05 & 0 \\
$e$						& 0.001 & $\delta \lambda$ (km) & -60 & -50 & -40 \\
$i \ (\degree)$			& 91 & $\delta e_x$ (km) 		& -0.05 & 0.5 & 0.15 \\
$\Omega \ (\degree)$ 	& 0 & $\delta e_y$ (km) 		& -0.1 & 0.3 & 0 \\
$\omega \ (\degree)$ 	& 0 & $\delta i_x$ (km) 		& -0.5 & -0.45 & -0.15 \\
$M_0 \ (\degree)$ 		& 0 & $\delta i_y$ (km) 		& 0.05 & 0.1 & 0.2\\
\hline
\end{tabular}
\end{table}
\par
A typical CubeSat star tracker produces bearing angle measurement noise on the order of 20 arcsec ($1\sigma$) \cite{bct}. Applying Equation \ref{spdistort} with $\delta \lambda = 100$ km, short-period distortions of almost 500 arcsec can be observed in extreme cases, which is well above star tracker noise. However, if formations are constrained to ROE magnitude ratios of $\delta \lambda / \delta e \geq 20$ and $\delta \lambda / \delta i \geq 20$ (i.e. so that along-track separations are dominant) for $10 \leq \delta \lambda \leq 200$ km, the maximum difference in short-period distortions between targets is $\leq10$ arcsec. This indicates that remaining errors after applying the transform are below expected $1\sigma$ noise. 
As an added benefit, targets tracks which appear very similar in $\mathcal{T}$ can become more well-separated and distinguishable in targets' differential frames. Furthermore, in the case of measurement assignment errors $-$ for example, if measurements are swapped between Targets $i$ and $j$ $-$ the total error in $\mathcal{T}_{i/j}$ is the sum of both assignment errors, since errors now affect both the frame origin and the track. Errors become easier to distinguish in such cases. 
\subsection{Measurement Prediction}
\label{prediction}
\noindent After applying the above transform, the only quickly-varying term on the right hand side of Equation \ref{VBSmotion} is observer true anomaly $f_o$ (or the equivalent term, depending on orbit regime). This provides expectations which can be leveraged, in that even if specific ROE are unknown, target motion must still be parametric with known form.
\par
First, the parameter $f_k$ which generated each track measurement $(\alpha, \epsilon)_k$ is obtainable from the observer's absolute orbit estimate, as are other relevant OE. This allows the model of Equation \ref{VBSmotion} to be simply and easily fitted to hypothesised tracks, and the resulting fit can be used to predict future track measurements. For bearing angles defined in the RTN frame, azimuth corresponds to $\hat{\bm{r}}_z^\mathcal{R}$ and elevation to $\hat{\bm{r}}_x^\mathcal{R}$. Thus, parametric target motion in bearing angle space can be written as
\begin{align}
\renewcommand{\arraystretch}{1.2}
\begin{bmatrix}
\epsilon\\
\alpha
\end{bmatrix}^\mathcal{R}
& \mkern-16mu \approx
\frac{r}{a}
\begin{bmatrix}
x_1 - x_2 (c_{f - x_3} + \frac{e}{2} c_{2f - x_3})\\
x_4 + x_5 s_{f + \omega - x_6}
\end{bmatrix}
\label{fit}
\\
& \mkern-16mu =
\frac{r}{a}
\begin{bmatrix}
x_1 \mkern-2mu - \mkern-2mu x_2 s_{x_3} (s_f \mkern-2mu + \mkern-2mu \frac{e}{2} s_{2f}) \mkern-2mu - \mkern-2mu x_2 c_{x_3} (c_f \mkern-2mu + \mkern-2mu \frac{e}{2} c_{2f}) \\
x_4 \mkern-2mu + \mkern-2mu x_5 c_{x_6} s_{f + \omega} \mkern-2mu - \mkern-2mu x_5 s_{x_6} c_{f + \omega}
\end{bmatrix}
\end{align}
\noindent where constants $\vec{x} = (x_1,...,x_6)^\top$ are scaled bearing angle equivalents of the EROE in Equation \ref{VBSmotion}. Given a set of $(\alpha, \epsilon)_k$ measurements and their respective $f_{k}, r_{k}, a_{k},\omega_{k}$ for $k = 1, ..., n$, $\vec{x}$ can be estimated by solving a pair of separable linear systems in elevation and azimuth, via
\begin{align}
\setlength\arraycolsep{1pt}
\renewcommand{\arraystretch}{1.3}
\begin{bmatrix}
\frac{r_1}{a_1} (c_{f_1} \mkern-4mu + \mkern-4mu \frac{e_1}{2} c_{2 f_1}) & \frac{r_1}{a_1} (s_{f_1} \mkern-4mu + \mkern-4mu \frac{e_1}{2} s_{2 f_1}) & \frac{r_1}{a_1} \\
\vdots & \vdots & \vdots \\
\frac{r_n}{a_n} (c_{f_n} \mkern-4mu + \mkern-4mu \frac{e_n}{2} c_{2 f_n}) & \frac{r_n}{a_n} (s_{f_n} \mkern-4mu + \mkern-4mu \frac{e_n}{2} s_{2 f_n}) & \frac{r_n}{a_n}
\end{bmatrix}
\mkern-4mu
\begin{bmatrix}
y_1\\
y_2\\
y_3
\end{bmatrix}
\mkern-4mu
=
\mkern-4mu
\begin{bmatrix}
\epsilon_1\\
\vdots\\
\epsilon_n
\end{bmatrix}&
\label{ls1} \\
\begin{bmatrix}
\frac{r_1}{a_1} c_{f_1 + \omega_1} & \frac{r_1}{a_1} s_{f_1 +\omega_1} & \frac{r_1}{a_1} \\
\vdots & \vdots & \vdots \\
\frac{r_n}{a_n} c_{f_n + \omega_n} & \frac{r_n}{a_n} s_{f_n + \omega_n} & \frac{r_n}{a_n} \\
\end{bmatrix}
\mkern-4mu
\begin{bmatrix}
y_4\\
y_5\\
y_6
\end{bmatrix}
\mkern-4mu
=
\mkern-4mu
\begin{bmatrix}
\alpha_1\\
\vdots\\
\alpha_n
\end{bmatrix}&
\label{ls2}
\end{align}
\noindent where coefficients $\vec{y} = (y_1, ..., y_6)^\top$ are related to $\vec{x}$ by
\begin{alignat*}{3}
x_1 &= y_3 								& x_4 &= y_6 \\
x_2 &= \sqrt{y_1^2 + y_2^2} 			& x_5 &= \sqrt{y_4^2 + y_2^5}  \\
x_3 &= \textrm{atan2}(-y_2, -y_1) \quad	& x_6 &= \textrm{atan2}(-y_4, y_5)
\end{alignat*}
\noindent Equations \ref{ls1} - \ref{ls2} are written more compactly as $\bm{A}_1 \vec{y}_1 = \vec{\epsilon}$ and $\bm{A}_2 \vec{y}_2 = \vec{\alpha}$ respectively. Notably, only three measurements are required to define a solution, which is well-suited to slow VBS measurement rates. Typical least squares methods such as QR decomposition can be used to solve each system and recover a target motion model in bearing angle space. The expected $(\alpha, \epsilon)_\textrm{pred}$ in a new image is computed using Equation \ref{fit} and the observer absolute orbit estimate at that epoch.
\par
At the beginning of a new track segment (e.g. after a maneuver) there may be too few measurements to fit the model. If only one prior measurement exists, the next predicted bearing angle is simply the previous angle. If only two prior measurements exist, the predicted bearing angle is computed linearly via
\begin{equation}
    (\alpha, \epsilon)_\textrm{pred} \approx (\alpha, \epsilon)_k + [(\alpha, \epsilon)_k - (\alpha, \epsilon)_{k - 1}]
\end{equation}
\par
%
%
If a navigation filter is present and estimating targets' relative states, this is instead used to provide predicted measurements for SAMUS by 1) propagating filter states and covariances into the new image epoch, and then 2) performing an unscented transform from the filter state space into bearing angle space \cite{josh, generalized}.
\subsection{Measurement Gating}
\label{gating}
\noindent The formation of new tracks is gated such that tracks must remain physically reasonable according to the assumption of consistent parametric motion (within a track segment). SAMUS applies a set of kinematic rules to each possible measurement-to-track assignment and only those which pass all rules are kept. The kinematic rules are:
\begin{enumerate}[itemsep=-1mm]
\item Track velocities must be below a set maximum.
\item Track velocities must be consistent over time.
\item Tracks should generally not feature acute angles.
\item Tracks should turn in a consistent direction.
\item Assigned measurements must be close to the predicted measurement.
\end{enumerate}
\par
Rule 1 stems from the knowledge that in angles-only scenarios of interest \cite {marsao, starling, argon, avanti}, targets are in similar orbits to observers. The magnitude and velocity of target relative motion in the tracking frame depends on the ROE. Objects with dissimilar orbits have large ROE and proportionally large track velocities. Consequently, a velocity threshold can be placed on tracks depending on the maximum allowed swarm ROE, where for step $\vec{v}_k$ it is required that $d_k < d_{\textrm{max}}$. As an example, for a near-circular orbit in LEO, $d_{\textrm{max}} \approx 0.005$ rad/min in bearing angle space allows $|\delta e / \delta \lambda| \leq 0.05$ and $|\delta i / \delta \lambda| \leq 0.05$. Note that eccentric orbits will observe larger $d_k$ near periapsis and higher-altitude orbits will observe comparatively lower $d_k$.
\par
Rule 2 denotes that track velocities should be relatively consistent between images. Velocities are only constant when $e \approx 0$ and target tracks are circular, which occurs if $\delta e = \delta i$ and $\delta \bm{e} \parallel \delta \bm{i}$. Otherwise, velocity variations grow with track aspect ratio $a_e/b_e$ (as per Equation \ref{ellipse3}) and orbit eccentricity. Two tests are applied, given by
\begin{align}
\frac{1}{r_{\textrm{max}}} \sum_{i=k-j}^{k-1} \frac{d_i}{j} &< d_k < r_{\textrm{max}} \sum_{i=k-j}^{k-1} \frac{d_i}{j}
\label{gating2} \\
\frac{1}{r_{\textrm{max}}} &< \frac{d_k}{d_{k-1}} < r_{\textrm{max}}
\label{gating3}
\end{align}
\noindent These tests imply that 1) the size of the new track step must be similar to the average size across the previous $j$ epochs, and that 2) ratios of successive step sizes must fall within an $r_{\textrm{max}}$ bound, defined by
\begin{equation}
r_{\textrm{max}} = \Big(1 + \frac{a_e}{2b_e} + \frac{10\sigma_{\textrm{VBS}}}{d_{\textrm{mean}}}\Big) \Big(1 + e_o\Big)
\end{equation}
\noindent where $\sigma_{\textrm{VBS}}$ is the bearing angle measurement noise of the VBS ($1\sigma$) and $d_{\textrm{mean}}$ is the mean of $d_{1,...k-1}$. Thus, $r_{\textrm{max}}$ has a minimum of 1.5 and allows larger velocity variations with larger track aspect ratios and/or orbit eccentricity. The $\sigma_{\textrm{VBS}}$ term allows for proportionally larger effects of measurement noise when track motion is very small. $10 \sigma_{\textrm{VBS}}$ is used as a limit to provide a $5\sigma$ `buffer' against the effects of noise on temporally adjacent observations in a track. 
\par
Rule 3 defines a minimum angle $\psi$ between successive steps $\vec{v}$. For $a_e/b_e \approx 1$ and $e_o \approx 0$, $\psi$ is obtuse with $\psi \approx \pi$. Otherwise, $\psi$ is most acute where $d_k$ is small and most obtuse where $d_k$ is large. To allow for these variations, this rule uses
\begin{equation}
\psi_k > \psi_{\textrm{min}}
\end{equation}
\begin{equation}
\psi_\textrm{min} = \min\Big(\frac{5\pi}{6}, \frac{5\pi}{6}\frac{d_k}{\max(d_{\textrm{mean}}, 10\sigma_{\textrm{VBS}})}\Big) (1 - e_o)
\label{gating4}
\end{equation}
\noindent Thus, if new $d_k$ is small compared to $d_\textrm{mean}$ or $10\sigma_{\textrm{VBS}}$, or the absolute orbit is more eccentric, the minimum allowed angle is more acute. The rule also ensures $\psi_\textrm{min} \leq 150$\degree, which relates to the maximum expected time interval between images. Larger intervals imply fewer track steps and thus more acute angles between track steps, as per the angle sizes of an $n$-sided regular convex polygon; 150\degree implies a maximum measurement interval of $T/12$.
\par
Rule 4 ensures that target tracks turn in an approximately consistent clockwise (or anticlockwise) direction, following the  expected parametric form. It is defined as
\begin{equation}
\textrm{sign}(\zeta_k - \zeta_{k-1}) = \textrm{sign}(\zeta_{k-1} - \zeta_{k-2})
\label{gating5}
\end{equation}
\noindent Eccentric orbits can observe multiple changes in direction as the first- and second-order components of relative motion constructively or destructively interact throughout the orbit. However, these direction changes occur on a timescale of $T/4$, meaning that on the shorter timescales between images, the rule generally holds. Measurement noise can also cause violations when track velocities are small, so direction changes are only used for gating if $|\pi - \psi_k| > \pi/10$ and $d_k > 10\sigma_{\textrm{VBS}}$.
\par
Rule 5 guarantees that new assigned measurements must lie within some error region around the track's predicted measurement. If a target state estimate with covariance is available, the error region in bearing angle space is obtained via an unscented transform and a user-specified $\sigma$-bound. If a state estimate is unavailable, the error region $E$ for new measurement $\vec{r}_{\textrm{meas},k}$ is a circle centered on predicted measurement $\vec{r}_{\textrm{pred},k}$, with radius
\begin{equation}
r_E = \max(10\sigma_{\textrm{VBS}}, 2 d_{\textrm{mean}}) (1+e_o)
\label{gating6}
\end{equation}
\noindent A modified prediction must be applied if SAMUS observes a maneuver or a measurement gap, because the next measurement will be affected by either the maneuver (Equation \ref{control}) or secular ROE drift (Equations \ref{secdistort1} - \ref{secdistort}). However, these ROE-based equations cannot quantitatively predict changes in track motion models in bearing angle space (unless ROE are provided, e.g. by a navigation filter). As described, the fitted coefficients of track motion models in bearing angle space do not correspond exactly to ROE. Bearing angles do not provide range information, and as per Equation \ref{bearingangles}, bearing angles and the derived coefficients are effectively normalized by target range. Nevertheless, it is still possible to predict qualitative effects on the next measurement and form a modified $E$ for Rule 5.
\par
\begin{figure}[ht]
\centering
\includegraphics[width=0.9\columnwidth]{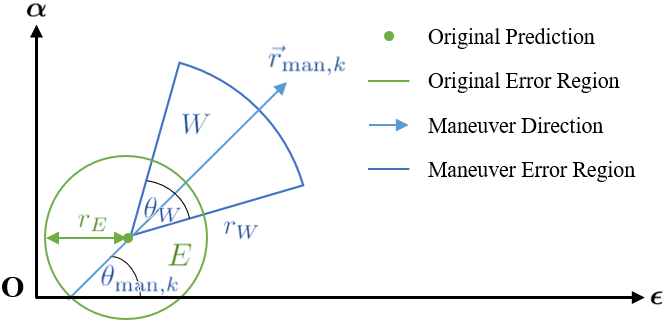}
\caption{Notional illustration of modified maneuver error region.}
\label{modified}
\end{figure}

For consistency, it is desired to view all swarm maneuvers as a target maneuvering relative to the observer; thus, if the expected maneuver is conducted by the observer, its components are first negated. RTN maneuver components are then rotated into the tracking frame to form $\delta \bm{v}^{\mathcal{T}}$. Then, as per Section \ref{frames}, $\delta v_x^{\mathcal{T}}$ describes a change in target relative velocity in the $\hat{\bm{x}}^{\mathcal{T}}$ or elevation direction, and $\delta v_y^{\mathcal{T}}$ determines a change in target relative velocity in the $\hat{\bm{y}}^{\mathcal{T}}$ or azimuth direction. When integrated over the time interval between images, a change in relative velocity leads to change in expected relative position (and expected measurement) in the new image. The post-maneuver predicted measurement $\vec{r}_{\textrm{man},k}$ is then located in some direction $\vec{v}_{\textrm{man},k}$ from the original prediction $\vec{r}_{\textrm{pred},k}$. The phase in the bearing angle plane that defines $\vec{v}_{\textrm{man},k}$ is $\theta_{\textrm{man},k} = \textrm{atan2}(\delta v_y^{\mathcal{T}}, \delta v_x^{\mathcal{T}})$. Figure \ref{modified} provides an illustration. The modified error region is the union of a 2D wedge $W$ $-$ defined by $\theta_{\textrm{man},k}$ and $\vec{r}_{\textrm{pred},k}$ $-$ and the original error region. Here, the radius of the wedge $r_W$ is defined as 20\% of the FOV with an arc angle $\theta_W = \pi/4$. The wedge must be large enough to ensure that true maneuvers are not excluded, but not too large as to allow formation of many false hypotheses. When instead resuming tracking after a measurement gap, the radius of the error region is doubled but its shape is unchanged, because secular ROE changes over a single orbit period are generally small.
\par
%
%
%
%
\par
For effective track gating, it is recommended to choose a measurement interval large enough that $d_{\textrm{mean}} \gg \sigma_{\textrm{VBS}}$ (i.e. so that noise does not wrongly validate or invalidate a large number of tracks), and small enough that sufficient information is received for robust stepwise prediction and gating. The authors have found that between 30-60 VBS images per orbit provides a suitable compromise.
\subsection{Measurement-to-Track Association}
\noindent Once a list of valid measurement-to-track associations has been created, the algorithm forms a set of valid global hypotheses. Each hypothesis is scored to assess its potential for selection, propagation or deletion. SAMUS employs an additive track score stemming from the kinematic rules. A hypothesis score is the sum of scores of each constituent track in its possible transformed frames. Consider hypothesis $i$, which features $p$ tracks (indexed by $l$) and thus $p - 1$ possible transforms for each track (indexed by $m$ with $l \neq m$). The hypothesis is scored across $q$ timesteps (indexed by $k$). Thus, define $s_{i, j, k, l, m,}$ as scoring criterion $j$ of timestep $k$, for track $l$ transformed with respect to track $m$. The total score for criterion $j$ of hypothesis $i$ is
\begin{equation}
s_{i, j} = \sum_{k = 1}^q \sum_{l = 1}^p \sum_{m = 1}^{p - 1} s_{i,j,k,l,m}
\end{equation}
\noindent and the final score of hypothesis $i$, with its component scores $s_{i, j}$ normalized to lie within $[0, 1]$, is
\begin{equation}
s_i = \sum_j \frac{s_{i,j} - \min_{i}(s_{i,j})}{\max_{i}(s_{i,j}) - \min_{i}(s_{i,j})}
\end{equation}
\noindent Criteria $s_{i,j}$ assess how well a track matches kinematic expectations. For an epoch $t_k$, the $s_{i,j}$ are defined as
\begin{align}
s_{i,1} =& ||\bm{A}_1 \vec{y}_1 - \vec{\epsilon}||_2 + ||\bm{A}_2 \vec{y}_2 - \vec{\alpha}||_2 \\
s_{i,2} =& ||\vec{r}_{\textrm{pred},k} - \vec{r}_{\textrm{meas},k}||_2 \\ 
s_{i,3} =& |d_k - d_{\textrm{pred}, k}| \\
s_{i,4} =& |d_k - d_\textrm{mean}| \\
s_{i,5} =& |\zeta_k - \zeta_{\textrm{pred}, k}| \\
s_{i,6} =& |\psi_k - \psi_{\textrm{pred}, k}| \\
s_{i,7} =& |\psi_k - \psi_\textrm{mean}| \\
s_{i,8} =& |f_k - f_o| \\
s_{i,9} =& 1/|d_k| \\
s_{i,10} =& 1/|\psi_k|
\label{scoring2}
\end{align}
\noindent Above, $s_1$ describes the residuals from track model fitting; $s_2$ describes the distance between the predicted measurement and new measurement; $s_3$ and $s_4$ describe the difference between the track step size and its predicted and mean sizes; $s_5$ describes the difference between the new track step phase and its predicted phase; $s_6$ and $s_7$ describe the difference between new track step angle and its predicted and mean angles; $s_8$ describes the difference between the new measurement parameter $f_k$ (computed from the previously-fitted model) and its parameter from the observer's absolute orbit estimate, $f_o$; and $s_{9}$ and $s_{10}$ bias the track towards smaller steps and larger step angles. The best hypothesis has the smallest score.
\par
In contrast to many methods which apply a single Mahalanobis distance metric between the predicted and assigned measurement for scoring, SAMUS aims to be more robust. Depending on the ROE and measurement frequency, target motion across images is often comparable in magnitude to VBS noise and multiple targets may be in close proximity in the image plane. A single scoring metric is not robust in these cases. By using a larger set of metrics $s_{1, ..., 10}$, score consensus supports the correct choice over time, even if some criteria temporarily support incorrect hypotheses. When compared to scoring with Equation \ref{MHTscore}, SAMUS avoids needing probabilistic estimates of false alarm densities or target decay rates which are not easily obtainable for spacecraft.
\par
When no measurements can be assigned to a track $-$ e.g. if that target was not present in the image $-$ its predicted measurement is used as a placeholder for propagation into future epochs. SAMUS also gives assigned measurements an `ambiguity' flag in that only unambiguous measurements should be passed to a navigation filter, to minimize false positives. Consider the best hypothesis $h_1$ with score $s_1$ and second-best $h_2$ with score $s_2$. For $h_1$ to be considered unambiguous, $s_1 < C_1 s_2$ for some constant $0 < C_1 < 1$. If satisfied, measurements in $h_1$ which have been members of their target's best track for $\geq C_2$ timesteps are unambiguous. In this work, $C_1 = 0.5$ is applied to ensure $h_1$ is superior in at least twice as many criteria. $C_2 = 3$ is used as this is the fewest number of measurements necessary for track fitting. All hypotheses $h_i$ with scores $s_i < C_3$ are also propagated, in case that they were the true hypothesis. 
A maximum score threshold of $C_3 = \max(3, 3 s_1)$ is chosen as a balance between robustness and computation cost. If a track is not visible for $\geq 0.1T$ (and was expected to be visible), it is deleted.
\par
If an angles-only navigation filter is estimating the complete swarm state, Malahanobis distance \cite{mtt1} is also used to score assignments, as per
\begin{equation}
s_{i, 11} = \sqrt{(\vec{r}_{\textrm{meas}} - \vec{r}_{\textrm{pred}})^\top \Sigma^{-1} (\vec{r}_{\textrm{meas}} - \vec{r}_{\textrm{pred}})}
\end{equation}
\noindent Here, $\Sigma$ is the predicted measurement covariance. To be considered unambiguous, no other kinematically-valid $\vec{r}_{\textrm{meas}}$ may be contained within the $3\Sigma$ region around $\vec{r}_{\textrm{pred}}$.
%
\subsection{Maneuver-to-Track Association}
\label{algorithm6}
\noindent If the batches of target measurements provided by SAMUS are to be applied for state estimation, the presence of maneuvers in each track must also be determined. All maneuvers known to occur at a specific epoch must be assigned to tracks at that epoch and hypothesis compatibility and scoring must take maneuvers into account. SAMUS assumes that it receives knowledge of all maneuvers performed by the swarm (i.e. their execution times, magnitudes, and directions) but that there is no immediate way to match this to tracked targets. Instead, the kinematics of tracks over time must be leveraged to determine maneuver correspondence. 
\par
In the first epoch after a maneuver, all tracks are split into two branches: one without a maneuver (in case the maneuver is not assigned to that track) and one with a maneuver (which begins a new track segment). If the observer maneuvers or all targets maneuver, all tracks are affected and non-maneuver branches are deleted. Assignment of maneuvers to specific tracks is performed when four new images after the maneuver have been received, to ensure that enough kinematic information is available for a robust assessment. Within each hypothesis, each constituent track is scored via six criteria, and the maneuver is assigned to the best-scoring track. 
\par
Define $\vec{x}_{\textrm{pre}}$ as the track model before the maneuver and $\vec{x}_{\textrm{post}}$ as the track model after the maneuver. The observed change is $\Delta \vec{x}_{\textrm{meas}} = \vec{x}_{\textrm{post}} - \vec{x}_{\textrm{pre}}$ and a predicted change can be computed with $\Delta \vec{x}_{\textrm{pred}} = \bm{B}_{\textrm{eroe}} \delta \bm{v}^{\mathcal{R}}$. This $\Delta \vec{x}_{\textrm{pred}}$ is not quantitatively accurate, because as discussed, $\vec{x}$ represent scaled EROE and $\bm{B}_{\textrm{eroe}} \delta \bm{v}^{\mathcal{R}}$ applies to true EROE. However, qualitative aspects of $\Delta \vec{x}$ can still be employed to check for consistency between the predicted and measured case. First, $\Delta \vec{x}$ are normalized using
\begin{equation}
  \Delta \vec{x}^N = \Delta \vec{x} / \max_j|\Delta \vec{x}_j|
\end{equation}
where $j = 1,...,6$ indexes components of $\Delta \vec{x}$. Subsequently, maneuver hypothesis scores $s_i^m$ are computed as
\begin{align}
    s_1^m &= ||\Delta \vec{x}_{\textrm{pred}}^N - \Delta \vec{x}_{\textrm{meas}}^N||_2\\
    s_2^m &= |\underset{j}{\textrm{argmax}}(\Delta x_{\textrm{pred},j}^N) - \underset{j}{\textrm{argmax}}(\Delta x_{\textrm{meas},j}^N)|\\
    s_3^m &= \sum_j [\textrm{sign}(\Delta x_{\textrm{pred},j}^N) - \textrm{sign}(\Delta x_{\textrm{meas},j}^N)]\\
    s_4^m &= |\theta_{\textrm{man},k-3} - \angle(\vec{r}_{\textrm{meas},k-3} - \vec{r}_{\textrm{pred},k-3})|\\
    s_5^m &= 1 / ||\Delta \vec{x}_{\textrm{meas}}||_1\\
    s_6^m &= \sum_{m = 0}^3 \frac{1}{|\vec{r}_{\textrm{meas},k-m} - \vec{r}_{\textrm{pred},k-m}(\vec{x}_{\textrm{pre}})|}
\end{align}
\noindent where $k$ indexes the current image epoch such that the maneuver occurred between $k-3$ and $k-4$. The notation $\vec{r}_{\textrm{pred},k-m}(\vec{x}_{\textrm{pre}})$ indicates predictions performed using pre-maneuver model $\vec{x}_1$. The lowest-scoring track is assigned the maneuver. Above, $s_1^m$ tries to match predicted and measured changes in the motion model; $s_2^m$ checks whether the component of $\Delta \vec{x}_{\textrm{meas}}$ that observed the largest change was as expected; $s_3^m$ checks whether components of $\Delta \vec{x}_{\textrm{meas}}$ changed in the expected directions; $s_4^m$ checks whether the maneuver caused the expected discrepancy between original predicted measurement and assigned measurement; $s_5^m$ biases maneuver assignments towards tracks with large changes in the motion model; and $s_6^m$ biases maneuver assignments towards tracks in which pre-maneuver and post-maneuver models are inconsistent.
\par
There may be cases in which maneuver information does not correspond to any visible or tracked targets. To allow for this, maneuvers are not assigned to a track if motion does not change after the maneuver epoch, i.e. $||\vec{x}_{\textrm{post}} - \vec{x}_{\textrm{pre}}||_2 \approx 0$ within a user-specified threshold.
\par
Finally, it is important to address tracking of completely unknown target maneuvers, which may arise when tracking unidentified, uncooperative, or adversarial spacecraft. If maneuvers are small such that the kinematic rules are not violated, and targets remain well-separated in bearing angle space, SAMUS can in principle continue tracking due to the overall robustness of kinematic scoring. Future work will explore tracking involving unknown maneuvers.
%
\subsection{Track Maintenance}
%
%
\noindent To initialize new targets, SAMUS applies the DBSCAN algorithm \cite{dbscan} to find clusters of unidentified measurements in the most recent four images. A DBSCAN cluster requires $\geq$ $n_D$ points within some maximum radius $\epsilon_D$, while other points are treated as noise. SAMUS aims to form $n_D/4$ new targets from each cluster and the prior gating and scoring criteria are applied to intra-cluster tracks to compute the best cluster hypothesis. Initialization does not require knowing the number of targets in advance.
\par
A SAMUS target is considered `finalized' when the algorithm is cooperating with a navigation filter that is estimating that target's swarm state, and the filter has converged to steady-state. In this scenario, tracking is greatly simplified as the filter is relied on to provide high-quality measurement prediction, gating and scoring information. Converged state covariances allow finalized targets to collapse down into a single hypothesis for typical formation geometries and VBS noise. 
\par
\begin{figure}[ht]
\centering
\includegraphics[width=0.9\columnwidth]{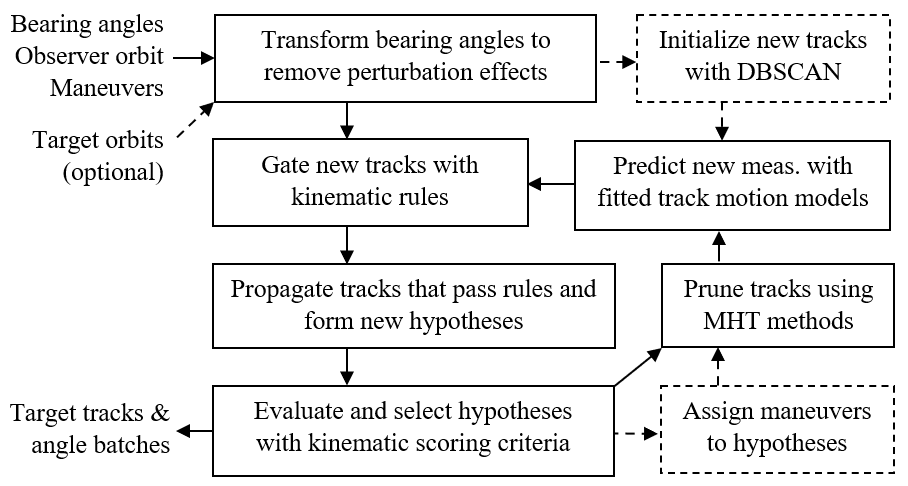}
\caption{SAMUS algorithm summary and core sequence of operations. Dashed lines denote steps that only occur at relevant epochs.}
\label{samuslogic}
\end{figure}

To more generally manage hypotheses, merge similar tracks, prune poor hypotheses, and limit computation costs, SAMUS employs several common methods as highlighted in Section \ref{mhtbg}. Algorithm \ref{trackmaintenance} in the appendix presents a pseudocode summary of these operations. Algorithm \ref{trackpropagation} in the appendix presents pseudocode for the main loop of SAMUS. Figure \ref{samuslogic} presents an overview of relevant operations and algorithms. When viewed as a whole, a potential weakness of SAMUS is the relatively large number of hyperparameters that affect performance. Although this paper provides suggested values for each, best practices for tuning will be addressed in future implementations
\section{Algorithm Validation}
\label{results}
\noindent Algorithm performance is validated across three suites of tests: 1) using synthetic bearing angle measurements, 2) using synthetic VBS imagery, and 3) using HIL imagery produced by a star tracker. SAMUS will also be flight-tested in 2022 aboard the NASA Starling mission as part of the Angles-only Real-time Trajectory Measurement System (ARTMS) software payload. ARTMS is a complete, autonomous, angles-only swarm navigation architecture for which SAMUS performs target identification and measurement assignment \cite{starling, artms}. Starling consists of four 6U CubeSats in LEO and its physical parameters form the basis for each simulation. Each Starling CubeSat will employ a Blue Canyon Technologies Nano Star Tracker (NST) for angles-only navigation. Intrinsic NST parameters were therefore applied when generating synthetic test images, listed in Table \ref{camparam}. An NST was also used to collect imagery for HIL tests. Tests were conducted on a PC with an Intel i7-7700HQ CPU and 16GB RAM.
\begin{table}[th]
\renewcommand{\arraystretch}{1.0}
\caption{Intrinsic parameters of the NST.}
\label{camparam}
\centering
\begin{tabular}{|l|l|}
\hline
Intrinsic Parameter & Value \\
\hline
Image Size (pixels) 		& 1280 $\times$ 1024 	\\
FOV	 ($\degree$)		& 12 $\times$ 10		\\
Pixel Size ($\mu$m)	& 5.3			\\
Focal Length (mm)	& 30.2			\\
Pixel Intensity Range & 0 - 255\\
\hline
\end{tabular}
\end{table}
\subsection{Data Generation}
\begin{figure}[th]
\centering
\includegraphics[width=0.9\columnwidth]{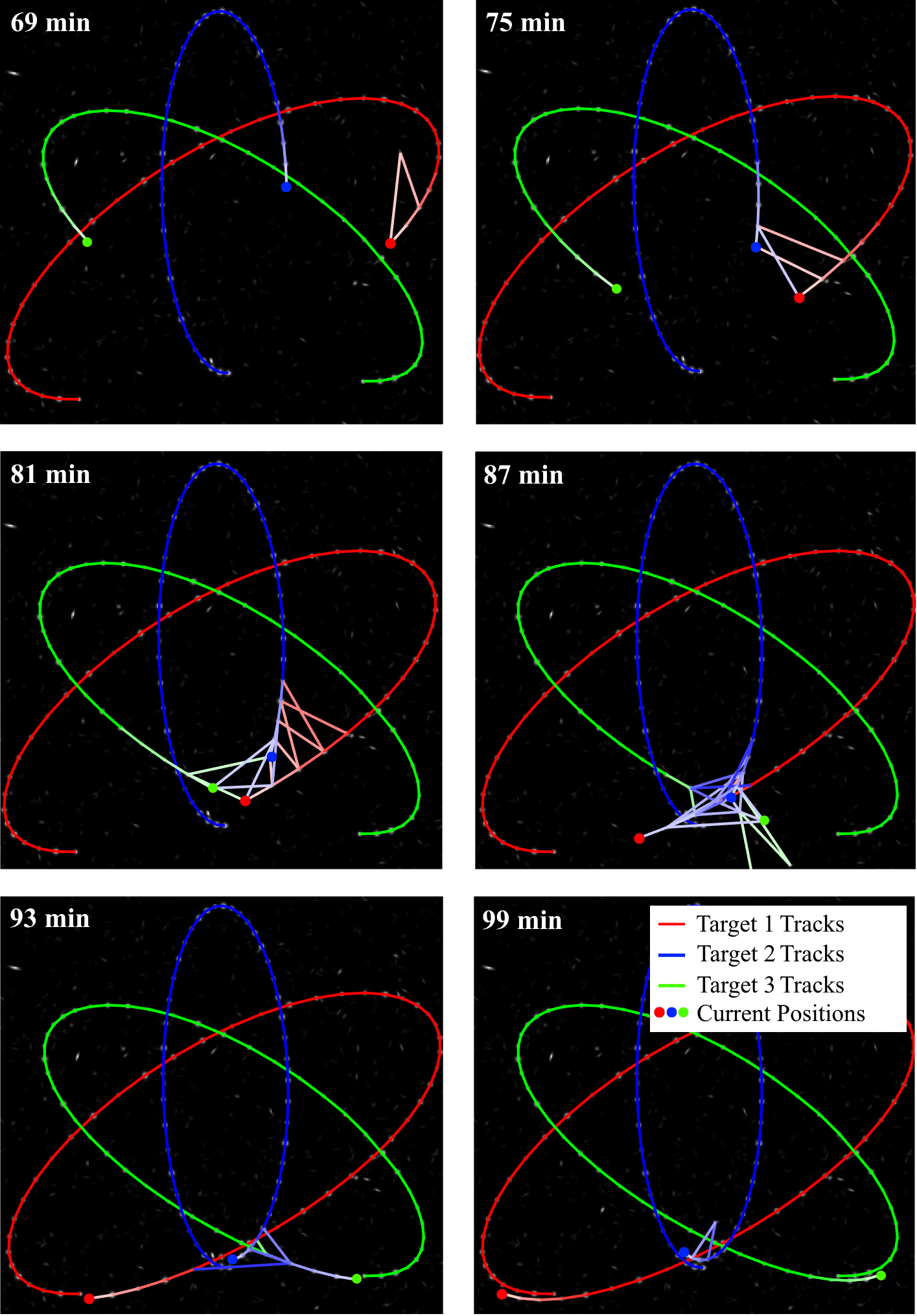}
\caption{Illustration of hypothesis evolution. Track hypotheses are overlaid on superimposed VBS images from $t = 0$ up to the specified time. Lighter track segments are newer and considered ambiguous while darker segments are more certain.}
\label{evotest}
\end{figure}
\begin{figure}[ht]
    \centering
    \includegraphics[width=\columnwidth, trim = {0.8cm 0.8cm 0.8cm 0.5cm}]{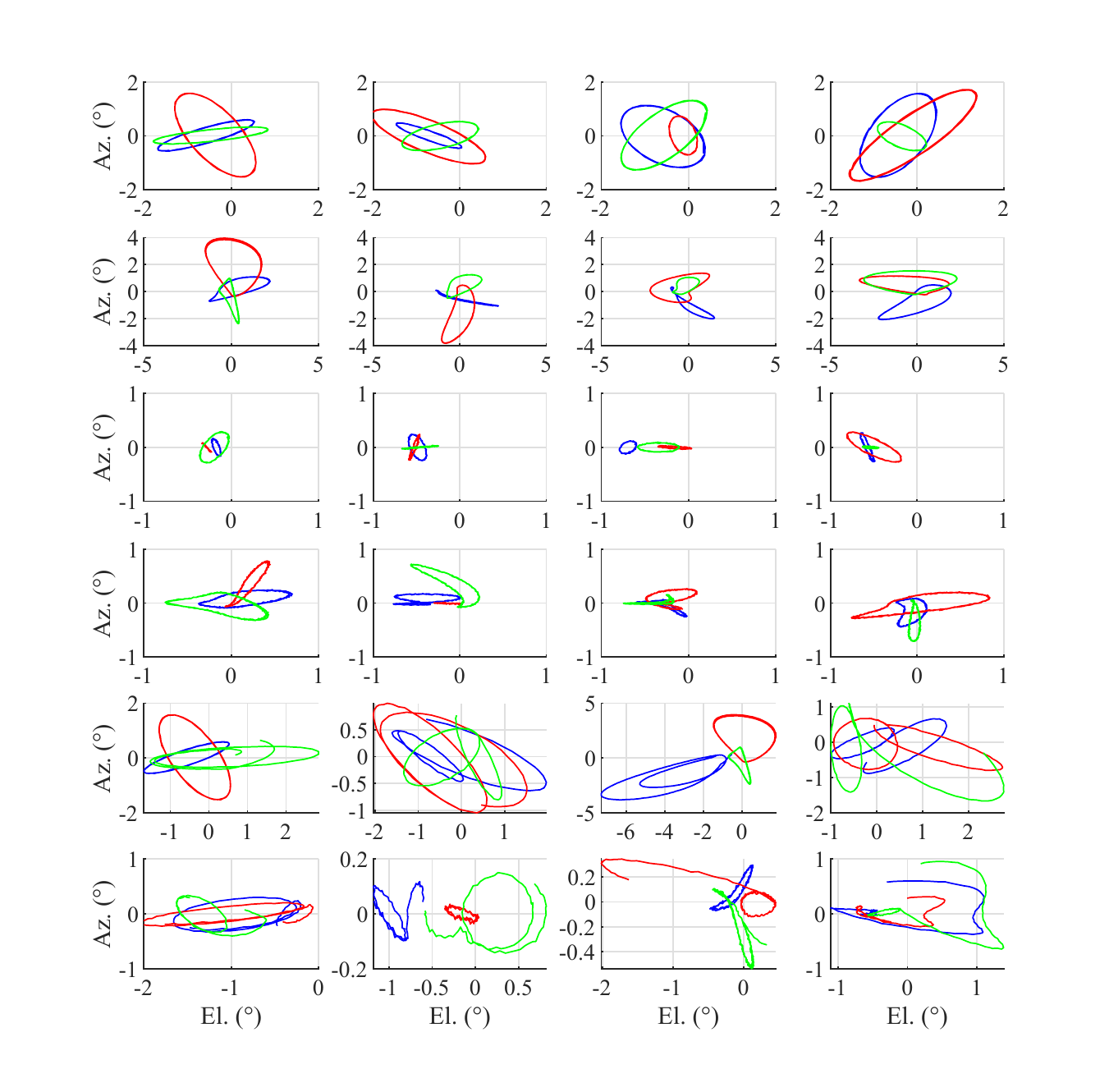}
    \caption{Examples of target motion in the VBS frame. Different colours represent distinct targets. Rows 1 - 6 correspond to datasets NC-EIS, ECC-EIS, NC-IT, ECC-IT, EIS-MAN and IT-MAN respectively.}
    \label{montecarlo}
\end{figure}
\noindent To generate test cases, the positions and velocities of a four-spacecraft swarm are numerically integrated using the Stanford Space Rendezvous Lab's $S^3$ software \cite{vince}. Perturbations include a 120x120 spherical harmonic gravity model, a cannonball drag model using Harris-Priester atmospheric density, a cannonball solar radiation pressure model with cylindrical Earth shadow, and third-body lunisolar gravity. Spacecraft attitude is fixed such that the VBS boresight points in the $\hat{\bm{y}}^\mathcal{W}$ direction. For tests using synthetic bearing angles, Gaussian white measurement noise is added to ground truth angles with zero mean and $\sigma_\textrm{VBS} = 20''$. Attitude noise is $\sigma_\textrm{off-axis} = 3''$ and $\sigma_\textrm{roll} = 20''$, which is considered typical for a modern CubeSat star tracker \cite{bct, centroid1} and verified in the HIL tests. Between 3 - 10 extra measurements are added to each bearing angle set to emulate passing satellites, debris and non-catalog stars, with positions pulled from a uniform distribution across the FOV. Observer absolute orbit knowledge is provided via a single ECI position/velocity estimate at the start of the simulation. This is propagated by SAMUS into subsequent epochs by numerical integration of the GVE, with a timestep of 30s and a 20x20 gravity model. No other perturbations are modeled by SAMUS. The inital estimate has Gaussian white noise of $\sigma_{\textrm{pos}} = 10$ m and $\sigma_{\textrm{vel}} = 0.02$ m/s. Maneuvers are executed with 5\% 1$\sigma$ magnitude error and 60'' 1$\sigma$ direction error \cite{manerror}.
\par
Each simulation consists of one observer tracking three targets for two orbits. Measurements are received every two minutes. Formations are generated from uniform distributions over the ranges of OE and ROE in Table \ref{results0}. To be consistent with earlier reasoning, limits of $\delta \lambda / \delta e \geq 20$ and $\delta \lambda / \delta i \geq 20$ are applied to ensure that targets remain in the FOV without any need for active camera tracking. Simulations are categorized by three aspects: type of absolute orbit, type of relative orbit, and inclusion of maneuvers. Absolute orbits are either near-circular (NC) with $e \in [0.0001, 0.01]$ or eccentric (ECC) with $e \in [0.01, 0.8]$. Relative orbits are either in-train (IT) or E/I-vector separated (EIS). In-train formations possess large differences in $\delta \lambda$ with other ROE being approximately zero, presenting a common but constrained case with little relative motion. EIS formations possess differences in all ROE and see more relative motion. Here, the limit that defines an IT formation is $\delta \lambda / \delta e \geq 200$ and $\delta \lambda / \delta i \geq 200$. Finally, some test cases include impulsive maneuvers (MAN) by the observer or targets. Maneuvers are generated via uniform distributions across spacecraft, execution time, direction, and magnitude, with $0.1 \leq |\delta \bm{v}^{\mathcal{R}}| \leq 2$ m/s. Two maneuvers occur in each simulation, if included. The complete simulation set consists of:
\begin{enumerate}[itemsep=-1mm]
    \item 200 NC-EIS and 100 NC-EIS-MAN
    \item 200 ECC-EIS and 100 ECC-EIS-MAN
    \item 100 NC-IT and 50 NC-IT-MAN
    \item 100 ECC-IT and 50 ECC-IT-MAN
\end{enumerate}
\par
\begin{table}[th]
\renewcommand{\arraystretch}{1.0}
\caption{OE and ROE ranges for simulations. $r_p$ is the radius of periapsis of the orbit.}
\label{results0}
\centering
\begin{tabular}{|l|l||l|l|}
\hline
Obs. OE & Range & Tar. ROE & Range\\
\hline
$r_p$ (km) 		&	[6750, 7150]			& $\delta a$ (km)		&	[-0.2, 0.2]	\\
$e$ 			&	[0.0001, 0.8]			& $\delta \lambda$ (km)	&	[5, 200]	\\
$i$				&	[0, $\pi$]			& $\delta e_x$ (km)		&	[-5, 5]	\\
$\Omega$		&	[0, $2\pi$]			& $\delta e_y$ (km)		&	[-5, 5]	\\
$\omega$		&	[0, $2\pi$]			& $\delta i_x$ (km)		&	[-5, 5]	\\
$M_0$			&	[0, $2\pi$]			& $\delta i_y$ (km)		&	[-5, 5]	\\
\hline
\end{tabular}
\end{table}
Figure \ref{evotest} presents an example of hypothesis evolution for three targets in an EIS formation in a near-cirular orbit. When targets are in close proximity, many new tracks are created; however, as more measurements arrive, tracks are scored and trimmed to converge to the true hypothesis. Figure \ref{montecarlo} presents examples of target tracks in simulations.
\subsection{Synthetic Bearing Angle Tests}
\begin{table*}[ht]
\renewcommand{\arraystretch}{1.0}
\caption{Monte-Carlo results $(1\sigma)$ for the the SAMUS algorithm for simulation subsets.}
\label{SAMUSresults}
\centering
\begin{tabular}{|l|l|l|l|l|l|l|l|}
\hline
Dataset & Precision (\%) & Recall (\%) & Accuracy (\%) & \begin{tabular}[c]{@{}l@{}}100\% Prec.\\ Cases (\%) \end{tabular} & \begin{tabular}[c]{@{}l@{}}Mean Error\\ (arcsec)\end{tabular} & \begin{tabular}[c]{@{}l@{}}Max. Error\\ (arcsec)\end{tabular} & \begin{tabular}[c]{@{}l@{}}Runtime per\\ Epoch (ms)\end{tabular} \\ \hline
NC  & $99.99 \pm 0.06$ & $97.12 \pm 2.09$ & $98.15 \pm 1.34$ &  98.0 & $25 \pm 1$ & 600 & $17 \pm 5$\\
ECC & $99.42 \pm 2.59$ & $95.47 \pm 8.25$ & $94.85 \pm 10.10$ & 86.0 & $30 \pm 22$ & 2200 & $18 \pm 6$\\
IT  & $99.20 \pm 3.05$ & $92.89 \pm 9.63$ & $91.99 \pm 11.75$ & 79.0 & $31 \pm 25$ & 2200 & $22 \pm 8$\\
EIS & $99.95 \pm 0.68$ & $97.88 \pm 1.64$ & $98.62 \pm 1.14$ &  98.5 & $26 \pm 8$ & 2200 & $15 \pm 3$\\
ALL & $99.71 \pm 1.83$ & $96.31 \pm 6.01$ & $96.54 \pm 7.31$ &  92.0 & $27 \pm 16$ & 2200 & $18 \pm 6$ \\
MAN & $99.62 \pm 2.14$ & $94.90 \pm 5.84$ & $95.45 \pm 9.67$ &  86.7 & $42 \pm 180$ & 12300 & $18 \pm 6$ \\ \hline
\end{tabular}
\end{table*}
\begin{table*}[ht]
\renewcommand{\arraystretch}{1.0}
\caption{Monte-Carlo results $(1\sigma)$ for common algorithms across the complete simulation set.}
\label{sotaresults}
\centering
\begin{tabular}{|l|l|l|l|l|l|l|l|}
\hline
Algorithm & Precision (\%) & Recall (\%) & Accuracy (\%) & \begin{tabular}[c]{@{}l@{}}100\% Prec.\\ Cases (\%) \end{tabular} & \begin{tabular}[c]{@{}l@{}}Mean Error\\ (arcsec)\end{tabular} & \begin{tabular}[c]{@{}l@{}}Max. Error\\ (arcsec)\end{tabular} & \begin{tabular}[c]{@{}l@{}}Runtime per\\ Epoch (ms)\end{tabular} \\ \hline
SAMUS   & $99.68 \pm 1.93$ & $95.86 \pm 5.98$ & $96.19 \pm 8.15$ &  90.2 & $32 \pm 104$ & 12300 & $18 \pm 6$\\
GNN     & $80.66 \pm 15.30$ & $87.83 \pm 3.76$ & $89.78 \pm 4.99$ & 37.0 & $434 \pm 665$ & 23400 & $30 \pm 42$\\
JPDA    & $78.79 \pm 13.96$ & $71.13 \pm 7.54$ & $88.05 \pm 3.54$ & 47.5 & $168 \pm 176$ & 18000 & $90 \pm 264$\\
MHT     & $82.15 \pm 10.36$ & $77.21 \pm 8.51$ & $88.52 \pm 4.45$ &  25.7 & $439 \pm 475$ & 18500 & $163 \pm 231$\\
PHD     & $89.97 \pm 11.12$ & $55.28 \pm 18.92$ & $81.98 \pm 10.28$ &  58.3 & $196 \pm 232$ & 10400 & $50 \pm 21$ \\ \hline
\end{tabular}
\end{table*}
\begin{table}[ht]
\renewcommand{\arraystretch}{1.0}
\caption{Monte-Carlo results for SAMUS across all datasets when varying simulation parameters. Default parameters are bolded.}
\label{results3}
\centering
\begin{tabular}{|l|l|l|l|}
\hline
\begin{tabular}[c]{@{}l@{}} Measurement noise \\ std. dev. \end{tabular}    & \begin{tabular}[c]{@{}l@{}}Precision\\ (\%)\end{tabular} & \begin{tabular}[c]{@{}l@{}}Recall\\ (\%)\end{tabular} & \begin{tabular}[c]{@{}l@{}}Accuracy\\ (\%)\end{tabular} \\ \hline
10''  & 99.81 & 97.66 & 97.05 \\
\textbf{20''}  & 99.68 & 95.86 & 96.19 \\
40''  & 91.21 & 83.24 & 88.73 \\\hline
\begin{tabular}[c]{@{}l@{}}Measurement gap\\ \end{tabular}    & \begin{tabular}[c]{@{}l@{}}Precision\\ (\%)\end{tabular} & \begin{tabular}[c]{@{}l@{}}Recall\\ (\%)\end{tabular} & \begin{tabular}[c]{@{}l@{}}Accuracy\\ (\%)\end{tabular} \\ \hline
\textbf{0\%}   & 99.68 & 95.86 & 96.19 \\
30\%  & 99.43 & 87.97 & 94.44 \\
60\%  & 99.31 & 74.91 & 92.56 \\\hline
\begin{tabular}[c]{@{}l@{}}Measurement\\ interval \end{tabular}    & \begin{tabular}[c]{@{}l@{}}Precision\\ (\%)\end{tabular} & \begin{tabular}[c]{@{}l@{}}Recall\\ (\%)\end{tabular} & \begin{tabular}[c]{@{}l@{}}Accuracy\\ (\%)\end{tabular} \\ \hline
1 min  & 92.73 & 95.34 & 94.07 \\
\textbf{2 min}  & 99.68 & 95.86 & 96.19 \\
4 min & 99.34 & 92.10 & 95.89 \\\hline \hline
\begin{tabular}[c]{@{}l@{}} Abs. orbit uncer- \\ tainty $(\sigma_{\textrm{pos}}, \sigma_{\textrm{vel}})$\end{tabular}    & \begin{tabular}[c]{@{}l@{}}Precision\\ (\%)\end{tabular} & \begin{tabular}[c]{@{}l@{}}Recall\\ (\%)\end{tabular} & \begin{tabular}[c]{@{}l@{}}Accuracy\\ (\%)\end{tabular} \\ \hline
\textbf{(10 m, 0.02 m/s)} & 99.68 & 95.86 & 96.19 \\
(1 km, 2 m/s)    & 99.55 & 94.74 & 94.77  \\
(10 km, 20 m/s)  & 99.45 & 93.63 & 93.57 \\\hline
\begin{tabular}[c]{@{}l@{}} Maneuver execution \\ error $(\sigma_{\textrm{mag}}, \sigma_{\textrm{dir}})$ \end{tabular}    & \begin{tabular}[c]{@{}l@{}}Precision\\ (\%)\end{tabular} & \begin{tabular}[c]{@{}l@{}}Recall\\ (\%)\end{tabular} & \begin{tabular}[c]{@{}l@{}}Accuracy\\ (\%)\end{tabular} \\ \hline
\textbf{(5\%, 60'')}   & 99.68 & 95.86 & 96.19 \\
(10\%, 120'') & 99.64 & 95.52 & 95.81 \\
(15\%, 180'') & 99.51 & 94.39 & 95.02 \\\hline
\begin{tabular}[c]{@{}l@{}}Initial $\delta \bm{x}_{\textrm{roe}}$ std. \\ dev. (as \% of $\delta\lambda$)\end{tabular}    & \begin{tabular}[c]{@{}l@{}}Precision\\ (\%)\end{tabular} & \begin{tabular}[c]{@{}l@{}}Recall\\ (\%)\end{tabular} & \begin{tabular}[c]{@{}l@{}}Accuracy\\ (\%)\end{tabular} \\ \hline
(0.1, 5, 0.1, ..., 0.1)     & 99.91 & 98.93 & 98.15 \\
(0.2, 10, 0.2, ..., 0.2)    & 99.80 & 95.25 & 96.40 \\
(0.4, 20, 0.4, ..., 0.4)    & 95.29 & 81.71 & 91.30 \\ \hline \hline
\begin{tabular}[c]{@{}l@{}}Measurement \\ input type \end{tabular}    & \begin{tabular}[c]{@{}l@{}}Precision\\ (\%)\end{tabular} & \begin{tabular}[c]{@{}l@{}}Recall\\ (\%)\end{tabular} & \begin{tabular}[c]{@{}l@{}}Accuracy\\ (\%)\end{tabular} \\ \hline
Synthetic angles    & 99.68 & 95.86 & 96.19 \\
Synthetic images    & 97.18 & 85.33 & 90.31 \\ \hline
\end{tabular}
\end{table}
Table \ref{SAMUSresults} presents Monte-Carlo results for SAMUS across the various simulation sets: near-circular, eccentric, in-train, and E/I-vector separated (without maneuvers); all simulations without maneuvers; and all simulations with maneuvers. Performance metrics of accuracy, precision and recall are computed using `true positives', or measurements correctly assigned to a target; `true negatives', or measurements correctly not assigned to a target; `false positives', or measurements incorrectly assigned to a target; and `false negatives', or measurements incorrectly unassigned to a target. Accuracy assesses overall performance, precision focuses on reliability of assignments, and recall focuses on frequency of assignments.
\begin{align}
\textrm{accuracy} &= \frac{\textrm{TP + TN}}{\textrm{TP + TN + FP + TN}}\\
\textrm{precision} &= \frac{\textrm{TP}}{\textrm{TP + FP}}\\
\textrm{recall} &= \frac{\textrm{TP}}{\textrm{TP + FN}}
\end{align}
\noindent Precision is considered the most vital metric because angles-only orbit determination filters are very sensitive to measurement errors \cite{josh} and a single false positive can cause degradation of the filter state estimate. Here, an assignment is defined as false positive if the assigned measurement was produced by a different target and is more than 5$\sigma_{\textrm{VBS}}$ from the ground truth measurement.
\par
Table \ref{sotaresults} presents a comparison with four other common MTT algorithms: GNN, JPDA, traditional MHT, and a PHD filter. Each is implemented within the MATLAB Sensor Fusion and Tracking Toolbox. The R2020a \cite{matlab} version was used to generate the results in this paper. The MATLAB algorithms were set up to perform MTT in bearing angle space using a dynamics model following Equation \ref{VBSmotion}, such that target states consisted of 2D bearing angle positions and velocities in the observer's tracking frame and measurements are of the target position.
\par
Examining the `ALL' row in Table \ref{SAMUSresults}, SAMUS assignment precision is 99.7\%, indicating that false positives are minimized as desired. Despite an emphasis on discarding ambiguous measurements, recall remains high at 96.3\% and sufficient data is retained for navigation. The metric of `100\% Precision Cases' examines the proportion of tests observing zero false positives. Simulations are promising in that perfect precision is achieved across the vast majority of formations and mean assignment error remains on the order of measurement noise. When comparing different simulation sets, it is clear that eccentric orbits prove more challenging and diminish performance slightly. More complex and unpredictable relative motion is observed at high eccentricities and which impacts the reliability of kinematic gating and scoring. In-train formations also see worsened performance because targets are much closer together in the image plane and noise becomes larger compared to track velocities and separations. The addition of maneuvers does decrease performance slightly, since these sudden changes to target trajectories introduce short periods of significant uncertainty. However, the majority of manuever cases are successfully tracked and maneuvers are assigned to targets with XX\% accuracy. Overall, SAMUS achieves the necessary performance in that precision is above 99\% across all datasets and excellent consistency and accuracy are displayed in varying conditions.
\par
In comparison to other algorithms in Table \ref{sotaresults}, SAMUS demonstrates visibly superior performance. Compared to the next-best algorithm (the PHD filter), it retains a 10\% precision advantage and a 30\% advantage in the number of simulations with perfect precision. Other algorithms also saw significantly more variable performance with noticeable degradation during more difficult eccentric or in-train scenarios, producing larger standard deviations in each metric. Mean assignment errors are also much higher, indicating a comparative lack of robustness. The $<$90\% precision displayed by GNN, MHT, JPDA and the PHD filter would likely make them unusable for angles-only navigation in orbit with any degree of reliability unless significant modifications were made. SAMUS also demonstrates an advantage in runtime in MATLAB, being an order of magnitude faster than traditional MHT and three times faster than the PHD filter. Although these runtimes are not optimized, this indicates that SAMUS should not exceed to computational costs of other common algorithms. Computational scalability versus swarm size will be quantitatively studied in future when SAMUS is implemented on a CubeSat flight processor.
\par
The first three sections of Table \ref{results3} presents SAMUS performance across the same simulation set as key parameters are modified: measurement noise, measurement availability, and measurement interval. Results largely follow expected trends. Increased noise leads to decreased performance because the reliability of the parametric motion model (used for track prediction, gating and scoring) is negatively affected. This is especially detrimental for in-train formations, for which $40''$ of noise can overwhelm target velocity between images, invalidating the kinematic rules and dramatically reducing accuracy. This similarly occurs when the measurement interval is halved. The time between images should then be chosen such that target velocity is larger than expected noise. Conversely, lower noise improves precision and recall, and faster measurements benefit EIS formations in particular since with more data, SAMUS can better predict subsequent behavior. VBS quality and onboard processing power are therefore important considerations for tracking, in addition to formation geometry. Reduced measurement availability impacts performance to a much lesser degree and the algorithm successfully handles long eclipse periods. The slightly lowered recall is due to periods of ambiguous tracking (e.g. when re-initializing tracks after an eclipse) becoming a greater proportion of the orbit. 
\par
Sections 4 - 5 of Table \ref{results3} discuss performance when quality of a-priori data is varied, in the form of the observer's absolute orbit estimate and discrepancy between observer maneuver knowledge and executed maneuvers. Clearly, SAMUS is robust to low-quality absolute orbit information. The absolute orbit is primarily used to compute rotations between frames and to provide data for track model fitting. Relatively large state errors have minimal effect on fitting accuracy compared to (for example) sensor noise, and a poor initial estimate still provides consistent motion when propagated. Similarly, maneuver execution errors have minor effects on performance and for the three levels of error, maneuvers are assigned with 93-95\% accuracy. SAMUS operates by examining the general change in track shape from a maneuver, and the simulated errors are generally not large enough to completely change maneuver outcomes.
\par
Usage of a-priori relative orbit knowledge is also investigated, in the form of cooperation with a navigation filter. In this scenario, SAMUS uses the filter state estimate to better identify and assign measurements to targets, and the filter subsequently employs SAMUS measurement assignments to update its state. The angles-only unscented Kalman filter developed by Sullivan was applied \cite{josh, generalized}, which uses bearing angle measurements to estimate target ROE. Section 6 of Table \ref{results3} presents results when varying the quality of a-priori state information, i.e. the filter's initial ROE state covariance. Target range, described by $\delta \lambda$, is the most weakly observable ROE and initial $\delta \lambda$ uncertainty is therefore dominant. An initial 5\% $\delta \lambda$ uncertainty significantly improves tracking, achieving near-perfect precision and recall. Conversely, an initial 20\% $\delta \lambda$ uncertainty negatively impacts tracking, particularly for in-train or intersecting formations in which the large state uncertainty leads to multiple valid measurement assignments. However, as long as the filter state estimate possesses $\leq$10\% range uncertainty, SAMUS can leverage this information for enhanced performance in challenging scenarios.
\subsection{Synthetic Image Tests}
The final row of Table \ref{results3} presents results when moving to synthetic input imagery. VBS images are generated using 3D vector graphics in OpenGL \cite{connor}. The visual magnitudes, angles, and proper motions of SO are obtained from the Hipparcos star catalog and any objects within the camera FOV are rendered using Gaussian point spread functions (PSF). Background noise is added to every pixel according to a uniform distribution with intensity $I \in [0, 10]$, producing centroiding errors of $\sim$0.1 pixels. SAMUS processes each image to generate input measurements for MTT. The Gaussian Grid algorithm was used for centroiding \cite{centroid1}, the Pyramid algorithm was used for star identification \cite{pyramid}, and the Q-method was used for attitude determination \cite{qmethod}.
\par
\begin{figure}[ht]
\centering
\includegraphics[width=0.95\columnwidth]{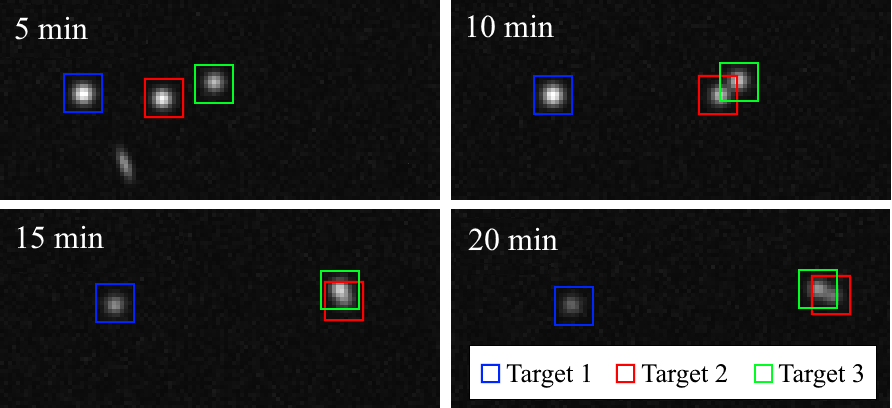}
\caption{Example of connected pixel clusters for an in-train formation (three targets).}
\label{connected}
\end{figure}

Moving to synthetic images decreases performance, especially for in-train formations, which specifically observed 6\% lower precision and 17\% lower recall. Overall, 29\% of IT formations and 8\% of EIS formations displayed at least one false positive. This reduction in performance stems from overlap of pixel clusters in simulated imagery $-$ frequently, the PSF of several objects will become connected (see Figure \ref{connected}). Traditional centroiding is unable to detect this and treats the joined PSF as one measurement, resulting in one `missing' measurement and one inaccurate measurement that is the average of the two. On occasion, a high overlap rate means SAMUS cannot distinguish targets or the joined measurement is inaccurate enough to be classified as an error. It is therefore useful to test for whether predicted measurements of separate targets are likely too similar to produce distinct pixel clusters, and for the algorithm to flag such assignments as ambiguous. Alternately, new centroiding techniques have explored detecting and separating joined PSF \cite{connect} and can be considered if primarily in-train tracking is desired.
\subsection{Hardware-in-the-Loop Tests}
\begin{figure*}[ht]
\centering
\includegraphics[width=6in, trim = {0cm 0cm 0cm 1cm}]{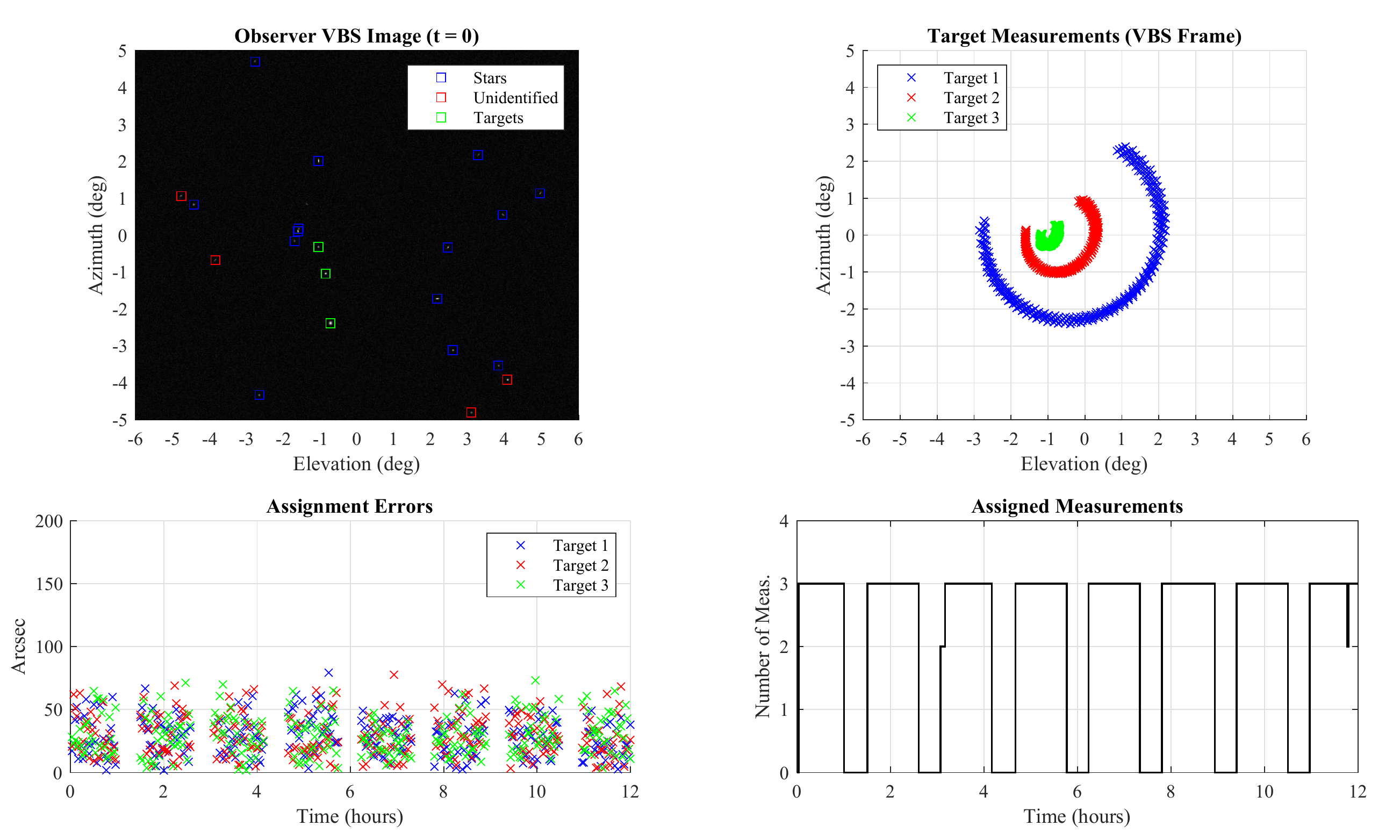}
\caption{Example HIL image (top left), target bearing angle measurements (top right), measurement assignment errors (bottom left) and measurement assignments (bottom right).}
\label{HITLtracks}
\end{figure*}

For HIL testing, input images are retrieved from a Blue Canyon Technologies NST as stimulated by the Stanford Space Rendezvous Lab's Optical Stimulator (OS). The OS is a a variable-magnification testbed consisting of two lenses and a microdisplay. A synthetic space scene is generated in accordance with previous sections (without background noise) and shown on the OS, and by moving the two lenses and display relative to each other, the NST is stimulated. The system is calibrated such that the VBS image is similar in both radiosity and geometry to what would be observed in orbit. Development, calibration and usage of the OS is detailed by Beierle et al. \cite{connor} with achievable errors between desired and measured bearing angles of less than $10''$. One such test is presented below, based on a proposed optical navigation experiment \cite{artms} for the aforementioned Starling mission. Formation OE and ROE are given in Table \ref{HITLparam}. Figure \ref{HITLtracks} presents the evolution of target tracks and bearing angle assignments across 12 hours or approximately 8 orbits. The observed assignment errors of up to $80''$ between assigned and ground truth measurements are reasonable when considering the expected error sources, which include NST error \cite{bct}, calibrated OS errors \cite{connor}, and attitude determination tolerances. Measurements are assigned to targets with 100\% precision and $>$98\% recall. This assignment performance is identical to the same test conducted with entirely synthetic imagery. SAMUS is therefore able to operate on representative camera images and flight scenarios.
\begin{table}[th]
\renewcommand{\arraystretch}{1.0}
\caption{Formation configuration for HIL test.}
\label{HITLparam}
\centering
\begin{tabular}{|c|c||c|c|c|c| }
\hline
OE & Obs. & ROE & Tar. 1 & Tar. 2 & Tar. 3 \\
\hline
$a$ (km) 				& 6978 & $\delta a$ (km) 		& 0 & 0 & 0 \\
$e_x$					& 0.0014 & $\delta \lambda$ (km)& 65 & 133 & 200 \\
$e_y$			        & 0.0014 & $\delta e_x$ (km) 	& 0 & 0 & 0 \\
$i \ (\degree)$ 	    & 98 & $\delta e_y$ (km) 		& 3 & 2.6 & 1.2 \\
$\Omega \ (\degree)$ 	& 40 & $\delta i_x$ (km) 		& 0 & 0 & 0 \\
$u \ (\degree)$ 		    & 105 & $\delta i_y$ (km) 	& 3 & 2.6 & 1.2\\
\hline
\end{tabular}
\end{table}
\section{Conclusion}
\label{conclusion}
This paper introduces the `Spacecraft Angles-only MUltitarget tracking System' (SAMUS) algorithm which is able to identify and track multiple target space objects from an observer spacecraft using only sequential images captured by the observer's on-board camera. The algorithm requires coarse absolute orbit knowledge of the observer but no knowledge of the target relative orbits, and provides multitarget measurement assignment capabilities as required for autonomous angles-only navigation of spacecraft swarms. Unlike prior flight experiments, which tracked single targets, and existing MTT methods, which possess limitations preventing their use aboard spacecraft, SAMUS takes advantage of relative orbit kinematics and multi-hypothesis techniques to consistently track multiple unknown targets. Tracking is performed for both near-circular and eccentric orbits and in the presence of partially-known swarm maneuvers.
\par
The underlying structure of SAMUS applies ideas from MHT, in that multiple simultaneous measurement assignment hypotheses are propagated with the aim of converging to the correct hypothesis over time. However, angles-only navigation in orbit necessitates extremely high assignment precision with low measurement frequencies and low computational costs. To make MHT suitable for this context, domain-specific knowledge and appropriate mathematical techniques are leveraged. First, it is observed that target tracks follow a parametric curve with known form in the RTN frame of the observer. Although perturbing forces such as $J_2$ gravity affect this form, the relative proximity of targets means they are affected similarly. Thus, by differencing target motion in the observer's reference frame, perturbations are cancelled between targets and parametric motion is recovered. Then, given sets of bearing angle measurements, motion models can be fitted to target tracks using linear system techniques. This allows simple, accurate prediction of target motion using as few as three prior measurements. From this model, a set of kinematic criteria is developed which target tracks must fulfil to be valid, and target tracks are scored by how well they match kinematic expectations. These track gating and scoring criteria enable effective trimming of unlikely hypotheses and simple, robust selection of likely hypotheses. Maneuvers are also be assigned to tracks by assessing subsequent changes in kinematic behavior.
\par
Monte-Carlo simulations display promising performance. For in-train and E/I-vector separated formations, SAMUS achieves 99.20\% and 99.95\% precision respectively and $>$92\% recall. It performs significantly better than other MTT algorithms, especially when the requirements for spaceborne angles-only navigation are taken into account, i.e. high precision, low computation cost and consistent performance across the formation parameter space. Performance is maintained under expected levels of measurement noise and with large measurement gaps. If an angles-only navigation filter is also estimating the swarm state, SAMUS can apply this information to achieve 100\% precision in difficult scenarios. Camera-in-the-loop tests verify performance under realistic conditions.
\par
SAMUS can also be straightforwardly extended to track spacecraft swarms at larger separations, or swarms in other dynamic environements such as lunar or cislunar space, by developing and implementing new parametric models for target motion that include the relevant dynamics and perturbation effects. For the case of low Earth orbit, work is currently underway to prepare a flight code implementation for the upcoming NASA Starling mission (2022). Starling intends to be the first ever demonstration of autonomous angles-only swarm navigation in orbit, for which SAMUS provides necessary multitarget measurement assignment capabilities as part of an on-board navigation software payload.
\bibliographystyle{elsarticle-num}   
\bibliography{bibliography}   

\begin{thebibliography}{10}
\expandafter\ifx\csname url\endcsname\relax
  \def\url#1{\texttt{#1}}\fi
\expandafter\ifx\csname urlprefix\endcsname\relax\def\urlprefix{URL }\fi
\expandafter\ifx\csname href\endcsname\relax
  \def\href#1#2{#2} \def\path#1{#1}\fi

\bibitem{intro1}
S.~D'Amico, M.~Pavone, S.~Saraf, A.~Alhussien, T.~S. Mohammed Al-Saud,
  S.~Buchman, R.~Byer, C.~Farhat, {Miniaturized Autonomous Distributed Space
  System for Future Science and Exploration}, in: 8th International Workshop on
  Satellite Constellations and Formation Flying, Delft, The Netherlands, 2015.

\bibitem{intro2}
O.~Brown, P.~Eremenko, {Fractionated Space Architectures: A Vision for
  Responsive Space}, in: Proceedings of the 4th Responsive Space Converence,
  Los Angeles, California, 2016.

\bibitem{intro4}
C.~Underwood, S.~Pellegrino, V.~J. Lappas, C.~P. Bridges, J.~Baker, {Using
  CubeSat/micro-satellite technology to demonstrate the Autonomous Assembly of
  a Reconfigurable Space Telescope (AAReST)}, Acta Astronautica 114 (2015)
  112--122.

\bibitem{marsao}
J.~Kruger, K.~Wallace, A.~W. Koenig, S.~D'Amico, {Autonomous Angles-Only
  Navigation for Spacecraft Swarms around Planetary Bodies}, in: 2021 IEEE
  Aerospace Conference, Big Sky, Montana, 2021.

\bibitem{starling}
H.~Sanchez, D.~McIntosh, H.~Cannon, C.~Pires, J.~Sullivan, S.~D'Amico,
  B.~O'Connor, {Starling1: Swarm Technology Demonstration}, in: 32nd Annual
  AIAA/USU Conference on Small Satellites, Logan, Utah, 2018.

\bibitem{argon}
S.~D'Amico, J.-S. Ardaens, G.~Gaias, H.~Benninghoff, B.~Schlepp, J.~L.
  Jørgensen, {Noncooperative Rendezvous Using Angles-Only Optical Navigation:
  System Design and Flight Results}, Journal of Guidance, Control, and Dynamics
  36~(6) (2013) 1576--1595.

\bibitem{avanti}
J.-S. Ardaens, G.~Gaias, {Noncooperative Rendezvous Using Angles-Only Optical
  Navigation: System Design and Flight Results}, Acta Astronautica 153 (2018)
  367--382.

\bibitem{intro3}
F.~Sellmaier, T.~Boge, J.~Spurmann, S.~Gully, T.~Rupp, F.~Huber, {On-Orbit
  Servicing Missions: Challenges and Solutions for Spacecraft Operations}, in:
  SpaceOps 2010 Conference, Huntsville, Alabama, 2010.

\bibitem{intro5}
B.~B. Reed, R.~C. Smith, B.~Naasz, J.~Pellegrino, C.~Bacon, {The Restore-L
  Servicing Mission}, in: 2016 AIAA Space Forum, Long Beach, California, 2016.

\bibitem{ao1}
D.~Woffinden, D.~Keller, {Optimal Orbital Rendezvous Maneuvering for
  Angles-Only Navigation}, Journal of Guidance, Control and Dynamics 32~(4)
  (2009) 1382--1387.

\bibitem{ao2}
G.~Gaias, S.~D'Amico, J.-S. Ardaens, {Angles-Only Navigation to a
  Noncooperative Satellite Using Relative Orbital Elements}, Journal of
  Guidance, Control, and Dynamics 37~(2) (2014) 439--451.

\bibitem{josh}
J.~Sullivan, S.~D'Amico, {Nonlinear Kalman Filtering for Improved Angles-Only
  Navigation Using Relative Orbital Elements}, Journal of Guidance, Control,
  and Dynamics 40~(9) (2017) 2183--2200.

\bibitem{generalized}
J.~Sullivan, A.~W. Koenig, J.~Kruger, S.~D'Amico, {Generalized Angles-Only
  Navigation Architecture for Autonomous Distributed Space Systems}, Journal of
  Guidance, Control and Dynamics (2021).

\bibitem{artms}
A.~W. Koenig, J.~Kruger, J.~Sullivan, S.~D'Amico, {ARTMS: Enabling Autonomous
  Distributed Angles-Only Orbit Estimation for Spacecraft Swarms}, in: 2021
  American Control Conference, New Orleans, Louisiana, 2021.

\bibitem{mtt0}
B.-n. Vo, M.~Mallick, Y.~Bar-shalom, S.~Coraluppi, R.~Osborne, R.~Mahler, B.-t.
  Vo, {Multitarget Tracking}, in: J.~G. Webster (Ed.), Wiley Encyclopedia of
  Electrical and Electronics Engineering, Springer International Publishing,
  2015.

\bibitem{mtt1}
S.~S. Blackman, R.~Popoli, {Design and Analysis of Modern Tracking Systems},
  Artech House, Boston, Massachusetts, 1999.

\bibitem{mtt4}
Y.~Bar-Shalom, P.~K. Willett, X.~Tian, {Tracking and Data Fusion: A Handbook of
  Algorithms}, YBS Publishing, Storrs, Connecticut, 2011.

\bibitem{mtt2}
R.~P.~S. Mahler, {Multitarget Bayes Filtering via First-Order Multitarget
  Moments}, IEEE Transactions on Aerospace and Electronic Systems 39~(4) (2003)
  1152--1178.

\bibitem{mtt3}
M.~Kisantal, S.~Sharma, T.~H. Park, D.~Izzo, M.~Märtens, S.~D'Amico,
  {Satellite Pose Estimation Challenge: Dataset, Competition Design and
  Results} (2020).

\bibitem{mtt5}
P.~Cano, J.~R. del Solar, {Robust Tracking of Soccer Robots Using Random Finite
  Sets}, IEEE Transactions on Intelligent Systems 32~(6) (2017) 22--29.

\bibitem{mtt6}
H.~Farazi, S.~Behnke, {Online Visual Robot Tracking and Identification using
  Deep LSTM Networks}, in: 2017 IEEE/RSJ International Conference on
  Intelligent Robots and Systems, Vancouver, Canada, 2017.

\bibitem{mtt7}
K.~LeGrand, K.~DeMars, {Relative multiple space object tracking using intensity
  filters}, in: Proceedings of the 18th International Conference on Information
  Fusion, Washington, DC, 2015.

\bibitem{vallado}
D.~A. Vallado, W.~D. McClain, {Fundamentals of Astrodynamics and Applications},
  4th Edition, Microcosm Press, Hawthorne, California, 2013.

\bibitem{maneuverfree}
J.~Sullivan, A.~Koenig, S.~D'Amico, {Improved Maneuver-Free Approach to
  Angles-Only Navigation for Space Rendezvous}, in: Proceedings of the 26th
  AAS/AIAA Space Flight Mechanics Conference, Napa, California, 2016.

\bibitem{damicothesis}
S.~D'Amico, {Autonomous Formation Flying in Low Earth Orbit}, Ph.D. thesis,
  Delft University, Delft, The Netherlands (2010).

\bibitem{nonsingular}
A.~Koenig, T.~Guffanti, S.~D'Amico, {New State Transition Matrices for Relative
  Motion of Spacecraft Formations in Perturbed Orbits}, Journal of Guidance,
  Control, and Dynamics 40~(7) (2017) 1749--1768.

\bibitem{linearised}
S.~D'Amico, O.~Montenbruck, {Proximity Operations of Formation-Flying
  Spacecraft Using an Eccentricity/Inclination Vector Separation}, Journal of
  Guidance, Control, and Dynamics 29~(3) (2006) 554--563.

\bibitem{michelle}
M.~Chernick, S.~D'Amico, {New Closed-Form Solutions for Optimal Impulsive
  Control of Spacecraft Relative Motion}, Journal of Guidance, Control, and
  Dynamics 41~(2) (2018) 301--319.

\bibitem{mht0}
S.~S. Blackman, {Multiple Hypothesis Tracking For Multiple Target Tracking},
  IEEE A\&E Systems Magazine 19~(1) (2004) 5--18.

\bibitem{mht1}
D.~Reid, {An Algorithm for Tracking Multiple Targets}, IEEE Transactions on
  Automatic Control 24~(6) (1979) 843--854.

\bibitem{mht2}
C.~Kim, F.~Li, A.~Ciptadi, J.~Rehg, {Multiple Hypothesis Tracking Revisited},
  in: 2015 IEEE International Conference on Computer Vision, Santiago, Chile,
  2015.

\bibitem{mht3}
I.~Cox, S.~Hongorani, {An efficient implementation of Reid’s multiple
  hypotheses tracking algorithm and its evaluation for the purposes of visual
  tracking}, IEEE Transactions on Pattern Analysis and Machine Intelligence
  18~(2) (1996) 138--150.

\bibitem{mht4}
S.~Deb, M.~Yeddanapudi, K.~Pattipati, Y.~Bar-Shalom, {A generalized S-D
  assignment algorithm for multisensor-multitarget state estimation}, IEEE
  Transactions on Aerospace and Electronic Systems 33~(2) (1997) 523--538.

\bibitem{mht5a}
R.~L. Streit, T.~E. Luginbuhl, {Maximum likelihood method for probabilistic
  multihypothesis tracking}, in: Proc. SPIE 2235, Signal and Data Processing of
  Small Targets, Orlando, Florida, 1994.

\bibitem{mht6}
S.~S. Blackman, {Multiple Target Tracking with Radar Applications}, Artech
  House, Dedham, Massachusetts, 1986.

\bibitem{mht7}
R.~W. Sittler, {An optimal data association problem in surveillance theory},
  IEEE Transactions on Military Electronics 8~(2) (1997) 125--139.

\bibitem{mht8}
K.~G. Murty, {An algorithm for ranking all the assignments in order of
  increasing cost}, Operations Research 16~(3) (1968) 682--687.

\bibitem{pyramid}
D.~Mortari, M.~Samaan, C.~Bruccoleri, J.~Junkins, {The Pyramid Star
  Identification Technique}, Journal of the Institute of Navigation 51~(3)
  (2004) 171--183.

\bibitem{bct}
D.~Hegel, {Small Spacecraft Subsystem State-of-the-Art: Attitude Determination
  and Control}, in: Proceedings of the 15th International Planetary Probe
  Workshop, The University of Colorado Boulder, 2018.

\bibitem{dbscan}
M.~Ester, H.-P. Kriegel, J.~Sander, X.~Xu, {A Density-Based Algorithm for
  Discovering Clusters in Large Spatial Databases with Noise}, in: Proceedings
  of the 2nd International Conference on Knowledge Discovery and Data Mining,
  AAAI Press, 1996, p. 226–231.

\bibitem{vince}
V.~Giralo, S.~D'Amico, {Development of the Stanford GNSS Navigation Testbed for
  Distributed Space Systems}, in: Proceedings of the 2018 International
  Technical Meeting of The Institute of Navigation, Reston, Virginia, 2018, pp.
  837--856.

\bibitem{centroid1}
T.~Delabie, J.~de~Schutter, B.~Vandenbussche, {An Accurate and Efficient
  Gaussian Fit Centroiding Algorithm for Star Trackers}, Journal of the
  Astronautical Sciences 61~(1) (1991) 60--84.

\bibitem{manerror}
E.~G. Lightsey, T.~Stevenson, M.~Sorgenfrei, {Development and Testing of a
  3-D-Printed Cold Gas Thruster for an Interplanetary CubeSat}, Proceedings of
  the IEEE 106~(3) (2018) 379--390.

\bibitem{matlab}
{MATLAB Statistics and Machine Learning Toolbox R2019a}, the MathWorks, Natick,
  Massachusetts, USA (2019).

\bibitem{connor}
C.~Beierle, S.~D'Amico, {High Fidelity Validation of Vision-Based Sensors and
  Algorithms for Spaceborne Navigation}, Journal of Spacecraft and Rockets
  56~(4) (2019) 1060--1072.

\bibitem{qmethod}
J.~R. Wertz, {Spacecraft Attitude Determination and Control}, 1st Edition,
  Springer Dordrecht, Dordrecht, The Netherlands, 1978.

\bibitem{connect}
W.~Ding, D.~Gong, Y.~Zhang, Y.~He, {Centroid estimation based on MSER detection
  and Gaussian Mixture Model}, in: 12th International Conference on Signal
  Processing, Hangzhou, China, 2014.

\end{thebibliography}
\appendix
\section*{Appendix: Algorithm Pseudocode}
\noindent Algorithm \ref{trackmaintenance} takes as input the set of all hypotheses which were propagated into the current epoch. It applies various criteria to trim unlikely hypotheses and merge similar hypotheses. The output is a reduced and more efficient set of hypotheses.
\par
\begin{algorithm}[ht]
\DontPrintSemicolon
\KwData{Propagated hypotheses}
\KwResult{Pruned hypotheses}
\If{$n_{\rm{targets}} > n_{\rm{targets, max}}$}{
	keep best $n_{\textrm{targets, max}}$ targets\;
}
get score $s_1$ of best hypothesis $h_1$\;
\For{$\rm{all\;hypotheses}$ $h_i$}{
	get $s_i$\;
	\If{$s_i > C_3$}{
		delete tracks existing in $h_i$ only\;
	}
}
\For{$\rm{all\;targets}$ $T_j$}{
	root node update at epoch $k - 8$\;
	\For{$\rm{all\;tracks}$ $t_m, t_n \in T_j$}{
		\If{$\rm{unobserved\;in}$ $\geq 10$\% $\rm{of\;visible\;period}$}{
			delete track\;
		}
		\If{$\rm{ambiguous\;for}$ $\geq 50$\% $\rm{of\;visible\;period}$}{
			delete track\;
		}
		\If{$t_m = t_n \: \forall$ $\rm{epochs}$ $k \in [0, 7]$}{
			keep best of $\{t_m, t_n\}$\;
		}
		\If{$t_m \neq t_n \: \forall$ $\rm{epochs}$ $k \in [0, 7]$}{
			keep best of $\{t_m, t_n\}$\;
		}
	}
	\If{$n_{\rm{tracks}} > n_{\rm{tracks, max}}$} {
		keep best $n_{\textrm{tracks, max}}$ tracks\;
	}
}
delete existing global hypotheses\;
cluster remaining tracks\;
re-form global hypotheses\;
\caption{Track maintenance algorithm.}
\label{trackmaintenance}
\end{algorithm}
Algorithm \ref{trackpropagation} takes as input the set of existing hypotheses and a new VBS image. It processes the image to obtain new bearing angle measurements; propagates existing tracks using the new measurements; and initializes new tracks. The output is an updated set of hypotheses and a chosen best hypothesis.
\begin{algorithm}[ht]
\DontPrintSemicolon
\KwData{Existing hypotheses and new image}
\KwResult{Updated best hypothesis}
get image from sensor\;
get absolute orbit estimate from satellite bus\;
perform image centroiding\;
perform star identification\;
perform attitude determination\;
rotate unidentified angles into tracking frame\;
\If{$\rm{relative\;orbit\;estimates\;exist}$}{
	compute predicted angles and covariances\;
	\If{$\rm{relative\;orbit\;estimates\;are\;new}$}{
		initialize new targets\;
	}
}
declare new list of propagated tracks $t_{\textrm{all}}$\;
\For{$\rm{existing\;tracks}$ $t_m$}{
	\For{$\rm{valid\;transforms}$ $t_n$}{
		fit motion model to track \;
		predict new measurement\;
		\For{$\rm{new\;measurements}$ $m_k$}{
			create new track $t_{\textrm{new}}$ from $t_m$ and $m_k$\;
			apply kinematic gating rules\;
			\If{$\rm{rules\;passed}$}{
				add $t_{\textrm{new}}$ to $t_{\textrm{all}}$\;
			}
		}
	}
}
form compatible hypotheses $h_i$ from $t_{\textrm{all}}$\;
\For{$\rm{hypotheses}$ $h_i$}{
	compute score $s_i$ using kinematic criteria\;
}
get score $s_1$ of best hypothesis $h_1$\;
initialize $n_{\textrm{hyp}}$ = 1\;
initialize new list of tracks to keep $t_{\textrm{keep}}$\;
\For{$\rm{hypotheses}$ $h_i$}{
	\If{$s_i < C_3$}{
		add $t_{\textrm{all}} \cap h_i$ to $t_{\textrm{keep}}$\;
		$n_{\textrm{hyp}} \mathrel{+{=}} 1$\;
	}
	\If{$n_{\textrm{hyp}} > 6$}{
		break\;
	}
}
\If{$\rm{maneuvers\;occurred\;at\;epoch}$ $k - 3$}{
    \For{$\rm{hypotheses}$ $h_i$}{
        score maneuver assignments for $t_m \in h_i$\;
        assign compatible maneuvers to $t_m \in h_i$\;
    }
}
\For{$\rm{targets}$ $T_j$}{
	\eIf{$\rm{measurement\;assigned}$}{
		store propagated tracks $t_{\textrm{keep}} \cap T_j$\;
	}{
		propagate $t_{\textrm{all}} \cap T_j$ from epoch $k - 1$ using predicted measurement\;
	}
	update ambiguity flags\;
}
run DBSCAN on remaining unidentified angles\;
\If{$\rm{clusters\;found}$}{
	apply kinematic gating rules\;
	initialize valid new targets\;
}
do track maintenance as per Algorithm \ref{trackmaintenance}\;
pass $h_1$ to output\;
\caption{Track propagation algorithm.}
\label{trackpropagation}
\end{algorithm}
\end{document}